\documentclass[a4paper,11pt]{article}
\pdfoutput=1 

\usepackage{jheppub} 
\usepackage[mathscr]{euscript}
\usepackage[T1]{fontenc} 

 \usepackage{musicography}
 
\usepackage{url}
\usepackage{bm}
\usepackage{verbatim}
 
 \usepackage{makecell}
 \usepackage{float}
 
 \usepackage[makeroom]{cancel}
 
 \newcommand{\bv}{ \begin{verbatim}}

\newcommand{\ket}[1]{\ensuremath{\left|#1\right\rangle}}

\newcommand{\be}{\begin{equation}}
\newcommand{\ee}{\end{equation}}
\newcommand{\bpm}{\begin{pmatrix}}
\newcommand{\epm}{\end{pmatrix}}

\newcommand{\PBK}[1]{\ensuremath{\begin{pmatrix}#1\end{pmatrix}}}

\newcommand{\beqn}{\begin{eqnarray}}
\newcommand{\eeqn}{\end{eqnarray}}

\usepackage{slashed}
\newcommand{\cD}{\mathcal D}

\newcommand{\cT}{\mathcal T}

\newcommand{\cO}{\mathcal O}
\newcommand{\cR}{\mathcal R}
\newcommand{\cP}{\mathcal P}
\newcommand{\cS}{\mathcal S}
\newcommand{\cW}{\mathcal W}

\newcommand{\cV}{\mathcal V}
\newcommand{\cZ}{\mathcal Z}

\newcommand{\cI}{\mathcal I}
\newcommand{\cN}{\mathcal N}
\newcommand{\cM}{\mathcal M}

\newcommand{\bH}{\mathbb H}
\newcommand{\bE}{\mathbb E}
\newcommand{\bC}{\mathbb C}
\newcommand{\bZ}{\mathbb Z}

 \usepackage{breqn}

\newcommand{\STr}{{\text{STr}}}

\newcommand{\vm}{{\text{vec}}}

\newcommand{\adj}{{\text{adj}}}

\newcommand{\p}{\partial}

 \newcommand{\GL}{{\rm GL}}
  \newcommand{\PSL}{{\rm PSL}}

\newcommand{\SYM}{\text{SYM}}

\usepackage{tikz}
\usetikzlibrary{arrows,shapes.misc}

\DeclareMathOperator{\Imag}{Im}
\DeclareMathOperator{\Tr}{Tr}

\DeclareMathOperator{\PE}{PE}

\DeclareMathOperator{\HB}{HB}

 \usepackage{tikz}

  \usepackage{mathtools}
\DeclarePairedDelimiter\floor{\Big\lfloor}{\Big\rfloor}
 
\author[\natural]{Hongliang Jiang }

 \affiliation[\natural{}]{Blackett Laboratory, Imperial College, London SW7 2AZ, U.K.}

 \emailAdd{h.jiang@imperial.ac.uk}

\title{\boldmath\huge  Modularity in Argyres-Douglas Theories \\ with $a=c$ }

\abstract{  We consider a family of Argyres-Douglas theories, which   are 4D $\mathcal N=2$ strongly coupled superconformal field theories (SCFTs)  but share many features with 4D $\cN=4 $ super-Yang-Mills theories. In particular, the  two   central charges of  these theories are the same, namely    $a=c$. 
We derive a simple and illuminating formula for the Schur index of  these theories, which factorizes into the product of a Casimir term and a term referred to as the Schur partition function. While the former is controlled by the anomaly, the latter   is identified with the vacuum character of the corresponding chiral algebra and is expected to satisfy the modular linear differential equation. 
Our  simple expression for the Schur partition function, which can be regarded as the generalization of MacMahon’s generalized sum-of-divisor  function, allows one to numerically compute the series expansions efficiently, and furthermore find the corresponding modular linear differential equation. In a special case where the chiral algebra is known, we are able to   derive the corresponding modular linear differential equation using Zhu's recursion relation. We further study the solutions to the modular linear differential equations and discuss their modular transformations. As an application, we    study the high temperature limit or the Cardy-like limit of the Schur index using   its simple expression and modular properties, thus shedding light on the 1/4-BPS microstates of   genuine  $\mathcal N=2$  SCFTs with $a=c$ and their dual quantum gravity via the AdS/CFT correspondence.

}

\begin{document} 
\maketitle
\flushbottom
\allowdisplaybreaks

\section{Introduction}

Supersymmetric and superconformal field theories have been attracting intensive interest  due to the feasibility  of  performing exact computations. Many protected sectors  have also been   discovered in supersymmetric and superconformal field theories,  which turn  out to enjoy  rich mathematical structures.  In the zoo of supersymmetric field theories, the 4D $\cN=2$ superconformal field theories are of particular  interest. On the one hand, the low energy limit on the Coulomb branch of 4D $\cN=2$ superconformal field theories (SCFTs), which is a   special K\"ahler geometry, can be effectively described using the famous Seiberg-Witten theory. On the other hand, the 4D $\cN=2$ superconformal field theories  also possess  an interesting protected sector, known as the Schur sector, which encodes the information about Higgs branch. 
The Schur sector is particularly interesting because there is a nice correspondence between the   Schur sector and  the mathematical notion of vertex operator algebra (VOA), dubbed SCFT/VOA correspondence \cite{Beem:2013sza}. One important item  in validating this correspondence is given by the so-called Schur index, which just counts the Schur operators in the SCFTs. In general, the Schur index has the following structures:

\be\label{SchurPS}
\cI  =q^{ c_{2d}/24} \cZ=1+\cdots~, \qquad \cZ=q^{ -c_{2d}/24} \cI=q^{ -c_{2d}/24} (1+\cdots)=\chi_{ \text{vac}}~.
\ee
To distinguish the  two quantities, we will refer to $\cI$  as Schur index, and  $ \cZ$ as Schur partition function. While the Schur index $\cI$ is more meaningful for counting and starts with 1 for the unique vacuum state,   it is the Schur partition function $\cZ$ that is identified with the vacuum character  $\chi_{ \text{vac} }$ of the corresponding VOA  following the SCFT/VOA correspondence \cite{Beem:2013sza}.  The Casimir factor $e^{ -c_{2d}/24} $ in $\cZ$ is crucial to ensure the nice modular behavior in the chiral algebra. \footnote{We will ues  vertex operator algebra and chiral algebra interchangably.}

The superconformal    index can be computed  in many theories using various techniques. Once the Schur index is known, the Schur partition function can be obtained trivially by multiplying the Casimir factor.  The multiplication  is straightforward but  appears somewhat artificial. It is natural to inquire whether the factorization of the Schur index \eqref{SchurPS}  can be achieved in a more natural manner.  We will demonstrate that the factorization of Schur index indeed naturally emerges in a family of Argyres-Douglas (AD) theories, which will be denoted as $\cT_{(p,N)}$ with two coprime integers $p=2,3,4,6$ and $N=2,3,4, \cdots$. 

 The Argyres-Douglas theories  $\cT_{(p,N)}$  can be obtained by conformally gauging several copies of AD theories $D_ {p_i}(SU(N))$ \cite{Cecotti:2012jx,Cecotti:2013lda} which have flavor symmetry $SU(N)$, where the set of $p_i$ depends on the value of $p$. The Argyres-Douglas theories  $\cT_{(p,N)}$ 
 share  many features with $\cN=4$ $SU(N)$ super-Yang-Mills (SYM) theories \cite{Buican:2020moo,Kang:2021lic}.  In particular, the two central charges are exactly the same $a=c$. More surprisingly, the Schur index of $\cT_{(p,N)}$   can be obtained from   the Schur index of $\cN=4$ $SU(N) $ SYM theory by specializing its fugacities to   particular values.  Meanwhile, the Schur index of $\cN=4$ $SU(N)$ SYM admits a closed formula when the flavor fugacities are turned off \cite{Bourdier:2015wda}. The generalization  to flavored Schur index  was established in \cite{Hatsuda:2022xdv}.  With the closed formula for the flavored Schur index of  $\cN=4$ $SU(N) $ SYM, we derived  the Schur index of $\cT_{(p,N)}$ in Eq.~\eqref{TpNSchur} \eqref{TpNZNN}. Surprisingly, the Schur index derived in this way is very simple and enjoys the  obvious and natural factorization  of the form in \eqref{SchurPS}. We are then led  to a remarkably simple formula for the Schur partition function  of  $\cT_{(p,N)}$ AD theory. 
 
Following the SCFT/VOA correspondence, the Schur partition function is identified with the vacuum character of the corresponding chiral algebra. In particular, this implies that Schur partition function satisfies the modular linear differential equation (MLDE) \cite{Beem:2017ooy}, $\cD_q^{(k)}\cZ=0$ where $\cD_q^{(k)}$ is the modular  linear differential operator of weight $2k$ (see Appendix~\ref{modformMLDE} for details). The   modular linear differential equations  transform  covariantly with specific weight under  modular 
transformations,  significantly constraining the structure of modular linear differential equations.
At a given weight, the modular linear differential equations  are almost fully determined, up to several constants. 
Such a kind of simplicity allows one     to numerically search for the modular linear differential equations  which  are satisfied by the Schur partition function. The numerical search becomes even simpler and tangible thanks to the simple closed form of the Schur partition function of $\cT_{(p,N)}$ AD theory.   
To illustrate, we will find the MLDEs in  several examples and study their solutions.

At a fundamental level, the  modular linear differential equation is rooted in the existence of a specific kind of null state in the corresponding VOA.  Once the null state is known, one can derive the modular linear differential equations systematically using   Zhu's recursion relation \cite{zhu} and the commutation relations in the chiral algebra. We will review the details in Appendix \ref{Zhurecursion}. Unfortunately, the chiral algebra of $\cT_{(p,N)}$ AD theory is generally complicated and not known explicitly, except for the  simplest case of $\cT_{(3,2)}$ AD theory. The VOA of $\cT_{(3,2)}$ AD theory is known \cite{Buican:2016arp,Feigin:2007sp}. With the explicit OPEs at hand, we manage  to   find the desired null  state, and derive the corresponding MLDE of weight 10. Such a MLDE can be   verified   numerically to  very high order.

As a physical application of MLDE, we further study the high temperature limit of the Schur index / Schur partition function. The high temperature limit of the Schur index of 4D $\cN\ge 2$ SCFTs has been studied previously in \cite{Eleftheriou:2022kkv,ArabiArdehali:2023bpq}  for $\cN=4$ SYM theories and some other $\cN=2$ SCFTs.
The  $\cT_{(p,N)}$  AD  theories of interest in this paper have    central charges $a=c$ and are honest 4D $\cN= 2$ SCFTs without enhanced supersymmetry. So understanding the high temperature limit of the Schur index of $\cT_{(p,N)}$  AD  theories is important and complementary to the examples studied before.
The modularity of solutions to MLDE enables one to derive the high temperature limit  systematically. Based on examples studied in this paper and some results in the literature, we are motivated to propose some conjectures on MLDE and particularly a power law asymptotic behavior for the Schur partition function in \eqref{ZTpNScaling}, which in the special case of $p=2$ can be proved using our simple formula for Schur partition functions.

The rest of the paper is organized as follows.  In section~\ref{TpNIndex}, we will  first introduce the   $\cT_{(p,N)}$ AD theories that will  be studied in this paper. Then we will derive a simple and illuminating formula for the Schur index  of $\cT_{(p,N)}$  which admits an obvious factorization   \eqref{SchurPS}.
In section~\ref{TpNMLDE}, we will study the modular linear differential equations in  $\cT_{(p,N)}$.    We will also discuss the solutions to the MLDEs and  their modular transformation behavior. 
In section~\ref{TpNHighT}, we will study the high temperature limit of the Schur index / partition function based on the MLDEs and modular properties. In section~\ref{conclusions}, we will summarize the main results of the paper and discuss several open questions for future explorations.  We also include a few technical appendices.
In appendix~\ref{modformMLDE}, we will review the concepts of  Eisenstein series and modular forms. We will also discuss the general structures of MLDEs and their solutions.  
In appendix \ref{Zhurecursion}, we review the basic concepts of VOA,   Zhu's recursion relation for torus one-point function, and discuss in detail how to derive MLDE from the null state of the VOA.
In appendix~\ref{MLDEAD}, we will present the explicit MLDEs in several families of AD theories, which are   either found  numerically or derived rigorously from the null state of the corresponding VOAs.

 \section{Schur partition function of $\cT_{(p,N)}$ Argyres-Douglas theory}\label{TpNIndex}
 
 In this section, we will first review the properties of $D_p(SU(N))$ AD theories, which are the building blocks of 
  $\cT_{(p,N)}$ AD theories.  After discussing the construction and  properties of  $\cT_{(p,N)}$ AD theories, we will then derive a simple formula for the Schur index of   $\cT_{(p,N)}$ AD theory.

 \subsection{$D_p(SU(N))$ AD theory}\label{DpSUNAD}
We will start with a specific type of 4D $\cN=2$ SCFTs denoted by  $D_p(G)$, which was introduced in \cite{Cecotti:2012jx,Cecotti:2013lda}.  In particular, in this paper we will be mainly focusing on the case of $G=SU(N)$, namely $D_p(SU(N))$ where $p,N>1$ are positive integers. Moreover, we will always impose the constraint    that  $p$ and $N$  are coprime. In this case, the flavor symmetry of $D_p(SU(N))$  is   $SU(N)$. \footnote{More precisely, the flavor symmetry is $PSU(N)$. However, in this paper, we will not be concerned about the  global form of the flavor symmetry, so we will just loosely say that the flavor symmetry is $SU(N)$ for simplicity. }

The $D_p(SU(N))$ theories are generalized  AD theories \cite{Argyres:1995jj,Xie:2012hs} and  have no direct $\cN=2$  Lagrangian  description.   
But they have  class $\cS$ realization: one can compactify the  6D (0,2)    SCFT  of type $A_{N-1}$  on a Riemann sphere with an irregular puncture and a full regular puncture. To manifest this construction,  one can equivalently use the following notation  \cite{Xie:2012hs,Song:2017oew}
\be
  D_{p}(SU(N))=(A_{N-1}^N [p-N],F)= (I_{N,p-N},F)~,
\ee
where $F$ means full puncture which is responsible for the $SU(N)$ flavor symmetry. The condition $\gcd(p,N)=1$ ensures that the  irregular puncture has no further contribution to flavor symmetry.
 
The flavor central charge $k_F$ and $c$ central charge are  \cite{Cecotti:2013lda}
\be
 k_F= \frac{2N(p-1)}{p}~,\qquad  c=\frac{(p-1) (N^2-1)}{12}~, 
\ee
while  $a$ central charge can be computed via
\be
8a-4c =\sum_{j=1}^{p-1}   \floor {\frac{N}{p} j}\Bigg(2\frac{N}{p} j- \floor {\frac{N}{p} j} \Bigg)  ~.
\ee

It turns out for our theories with $\gcd(p,N)=1$, there a simple formula for $a$ central charge: 
\be
\frac{a}{c}=1-\frac{1}{4p}~.
\ee
 
\subsection{$\cT_{(p,N)}$ AD theory}
Suppose we take a collection of $D_{p_i} (SU(N)))$ theory  and gauge the maximal diagonal flavor symmetry, the total one-loop $\beta$-function is  
\be
b=2\times N -   \sum_{i=1}^\ell  b_i
= \Big( 2  - \sum_{i=1}^\ell \frac{p_i-1}{ p_i } \Big) N~,
\ee
where the first contribution comes from the vector multiplet and the second contribution comes from the individual $D_{p_i} (SU(N))$ theory \cite{Cecotti:2012jx}. 

In order to get a  superconformal theory after gauging, we require that the beta function of gauge coupling should vanish,  $b=0$. This is equivalent to the requirement  that  $\sum_i k_{F }^{ D_{p_i}(SU(N))} =4 h^\vee (SU(N))=4N$.
Then one can easily show that the only possibilities are
\be\label{conformalcase}
  (p_1, p_2, \cdots, p_\ell )\quad=\quad(2,2,2,2)~, \qquad (3,3,3)~, \qquad (2, 4,4)~, \qquad (2, 3, 6)~.
\ee
 
 These conformally gauged AD  theories are the four series of theories that we will consider in this paper.  We will label them respectively as 
 \be\label{4theory}
\cT_{ (2,N)}~, \qquad \cT_{ ( 3,N)}~, \qquad  \cT_{ (4,N)}~,  \qquad  \cT_{ (6,N)}~.
 \ee 
 See figure~\ref{TpNfig} for the four series of theories. Therefore, the theories that we shall study    is generally denoted as $\cT_{(p,N)}$ where $p=2,3,4,6$ and $N=2,3,4\cdots$ subject to the condition that $\gcd(p,N)=1$.
  See \cite{Kang:2021lic} for further generalizations and    an alternative notation for this family of theories that we show in table~\ref{TpNss}.
     
A remarkable feature of these theories is that the $a $ and $c$ central charges are exactly the same. 
\beqn
c&=&\sum_i c_{{D_{p_i}}(SU(N))} +c_{\text{vec} } \times \dim  SU(N)   =\sum_i \frac{(p_i-1) (N^2-1)}{12}+\frac{1}{6}(N^2-1)~, 
\\
a&=&\sum_i a_{{D_{p_i}}(SU(N))} +a_{\text{vec} } \times \dim  SU(N)   =\sum_i \Big(1-\frac{1}{4p_i}\Big) \frac{(p_i-1) (N^2-1)}{12}+\frac{5}{24}(N^2-1)~. 
\eeqn
 Then it is easy to see that $a-c= b(N^2-1)/48=0$.
This property  is reminiscent of the  $\mathcal N=4$   SYM theories which have $a=c$ due to maximally superconformal symmetry.  
 
When $p=3,4,6$, we can also realize $\cT_{(p,N)}$ AD theory by considering IIB string compactified on isolated hypersurface  singularity (IHS), which is a Calabi-Yau three-fold   defined in terms of a  quasi-homogeneous polynomial  in $\bC^4$ \cite{DelZotto:2015rca}.  This geometric engineering way of constructing   $\cT_{(p,N)}$ AD theories   is very useful. For example, one can easily compute the Coulomb branch spectrum from the deformation of singularity. The Higgs branch dimension can also be obtained from the resolution of the singularity \cite{Closset:2021lwy}. By  considering small values of $N$, one can compute the Higgs branch dimension explicitly.~\footnote{This Higgs branch dimension is bit  tricky, as  the naive counting from $F$-term and  hyperkahler quotient  gives a zero dimensional Higgs branch. This turns out to be not true due to incomplete Higgsing for these $a=c$ theories:      on a generic point on the Higgs branch the gauge group is not completely broken and  there are still Coulomb directions in the full moduli space.
} The dependence  on $N$ turns out to be very simple and we find that the Higgs branch dimensions of   $\cT_{(p,N)}$  AD theories fit into the following simple formula
\be\label{TpNHBdim}
\dim_\bC \HB\Big(\cT_{(p,N)}\Big)=2\dim_\bH \HB\Big(\cT_{(p,N)}\Big)=2\floor{\frac {N }{ p}}~,
\ee
where $\floor{\bm\cdot}$ is the floor function. Note that for $p=2$, there is no known type IIB realization. However, it was argued  in \cite{Kang:2022zsl} that the        $\cT_{(2,2k+1)}$ AD theories have Higgs branches of  quaternionic dimension     $k$, which is consistent with \eqref{TpNHBdim}.

We list some properties of $\cT_{(p,N)}$ SCFTs in table~\ref{TpNss}. In particular, we find that the central charges are uniformlly given   by 
\be\label{TpNcentralcharge}
a=c=\Big( 1-\frac 1 p\Big) (N^2-1)~.
\ee

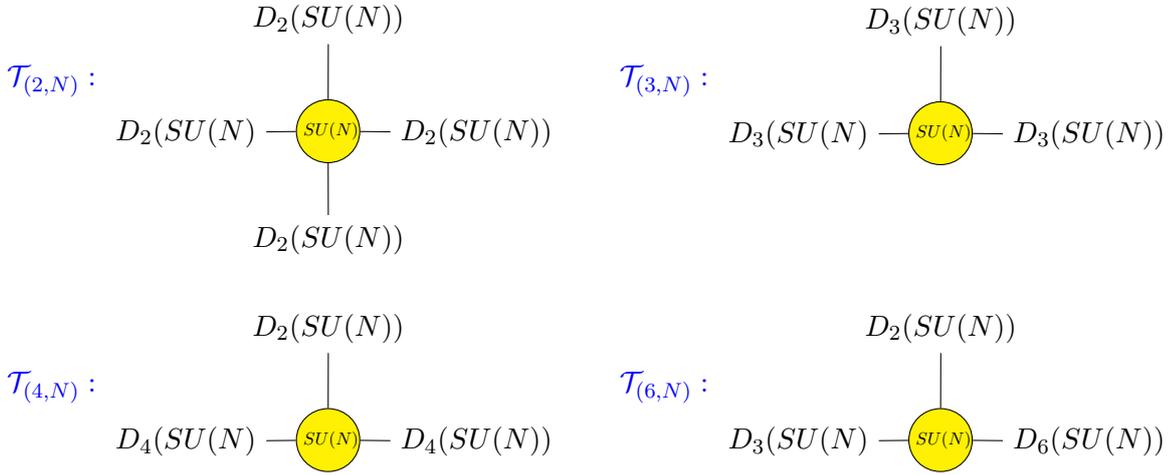
\begin{figure}
$\color{blue}\cT_{(2,N)}: $ \begin{tikzpicture}[baseline={([yshift= .591cm]current bounding box.center)}]
\node[anchor=south west](At) at (0.4,1.20) {$ {D}_{2}(SU(N))$};
\node[anchor=south west](A4) at (0.4,-1.720) {$ {D}_{2}(SU(N))$};
\node[anchor=south west](A1) at (-1.1-.3,-0.3) {$ {D}_{2}(SU(N)$};
\node[draw,fill=yellow,circle,minimum size=1mm,inner sep=-0pt](A2) at (1.53,0.0456781) {\scalebox{0.6}{ $  SU(N)$}};
\node[anchor=south west](A3) at (1.85+.5,-0.3) {$ {D}_{2}(SU(N))$};
\draw (A1)--(A2)--(A3);
\draw (At)--(A2)--(A4);
\end{tikzpicture}
\hspace{.5cm}
$\color{blue}\cT_{(3,N)}: $ \begin{tikzpicture}[baseline={([yshift=-.3ex]current bounding box.center)}]
\node[anchor=south west](At) at (0.4,1.20) {$ {D}_{3}(SU(N))$};
\node[anchor=south west](A1) at (-1.1-.3,-0.3) {$ {D}_{3}(SU(N)$};
\node[draw,fill=yellow,circle,minimum size=1mm,inner sep=-0pt](A2) at (1.53,0.0456781) {\scalebox{0.6}{ $  SU(N)$}};
\node[anchor=south west](A3) at (1.85+.5,-0.3) {$ {D}_{3}(SU(N))$};
\draw (A1)--(A2)--(A3);
\draw (At)--(A2);
\end{tikzpicture}
\hspace{.5cm}
\vspace{.5cm}

$\color{blue}\cT_{(4,N)}: $ 
 \begin{tikzpicture}[baseline={([yshift=-.3ex]current bounding box.center)}]
\node[anchor=south west](At) at (0.4,1.20) {$ {D}_{2}(SU(N))$};
\node[anchor=south west](A1) at (-1.1-.3,-0.3) {$ {D}_{4}(SU(N)$};
\node[draw,fill=yellow,circle,minimum size=1mm,inner sep=-0pt](A2) at (1.53,0.0456781) {\scalebox{0.6}{ $  SU(N)$}};
\node[anchor=south west](A3) at (1.85+.5,-0.3) {$ {D}_{4}(SU(N))$};
\draw (A1)--(A2)--(A3);
\draw (At)--(A2);
\end{tikzpicture}
\hspace{.5cm}
$\color{blue}\cT_{(6,N)}: $ 
 \begin{tikzpicture}[baseline={([yshift=-.3ex]current bounding box.center)}]
\node[anchor=south west](At) at (0.4,1.20) {$ {D}_{2}(SU(N))$};
\node[anchor=south west](A1) at (-1.1-.3,-0.3) {$ {D}_{3}(SU(N)$};
\node[draw,fill=yellow,circle,minimum size=1mm,inner sep=-0pt](A2) at (1.53,0.0456781) {\scalebox{0.6}{ $  SU(N)$}};
\node[anchor=south west](A3) at (1.85+.5,-0.3) {$ {D}_{6}(SU(N))$};
\draw (A1)--(A2)--(A3);
\draw (At)--(A2);
\end{tikzpicture}
\caption{$\cT_{(p,N)}$ from gauging of several copies of $D_{p_i}(SU(N))$. Yellow circle denotes $SU(N)$ gauge node. }
\label{TpNfig}
\end{figure}

\begin{table}[h] \centering\def\arraystretch{1.5}
\begin{tabular}{ |c|c|c|c|c| } 
\hline
 theory & $N$ & IIB string on IHS    & $a=c$ & $\dim_\bH$ HB \\
\hline
$\cT_{(2,N)}=\widehat D_4(SU(N))$ & $2k+1$ &  - & $\frac{1}{2} \left(N^2-1\right)$ &               $k$\\ 
$\cT_{(3,N)}=\widehat E_6(SU(N))$ & $3k+1, 3k+2$ &  $x^3+y^3+z^3 +w^N=0$   &     $\frac{2}{3} \left(N^2-1\right)$      & $k$ \\ 
$\cT_{(4,N)}=\widehat E_7(SU(N))$ & $2k+1$ &   $x^2+y^4+z^4 +w^N=0 $ &     $\frac{3}{4} \left(N^2-1\right)$      & $\floor{ \frac k 2}$\\ 
$\cT_{(6,N)}=\widehat E_8(SU(N))$ & $6k+1, 6k+5$ &  $ x^2+y^3+z^6 +w^N=0$ &   $\frac{5}{6} \left(N^2-1\right)$         & $k$\\ 
\hline
\end{tabular}
\caption{The  notations and properties of  $\cT_{(p,N)}$ theories. Here $N>1$ and $k,N \in \mathbb N$.  }
\label{TpNss}
\end{table}

Among the four infinite series of theories, there are several  special theories which admit other constructions. For example, we have 
\be
\cT_{(3,2)}=\widehat E_6(SU(2))=(A_2,D_4) 
=D_4^6[3]=D_4^4[2]~, 
\ee
where the last notation comes from the class $\cS$ realization by compactifying 6d (0,2) SCFT of type $D_4$ on a Riemann sphere with just one irregular puncture (and without any regular puncture) \cite{Xie:2012hs,Wang:2015mra}.  The notation $(A_2,D_4) $ also comes from the type IIB realization, which is different but equivalent to the type IIB realization in table~\ref{TpNss}.
Similarly, the other two special theories are
\be
\cT_{(6,5)}=\widehat E_8(SU(5))=(A_5,E_8)=E_8^{30}[6]~, \qquad\qquad  
\cT_{(4,3)}=\widehat E_7(SU(3))=(A_3,E_6)= E_6^{12}[4] ~.
\ee

\subsection{Schur partition function}

In 4D $\cN=2$ SCFTs,  
the Schur index is a useful quantity to  count  the  Schur operators which are $1/4$-BPS 
\be
\cI(q,\bm x)  =\Tr (-1)^F q^{E-R} \prod_j (x_j)^{F_j}~,
\ee
where the $E$ is  the energy,    $R$ is the $SU(2)_R$ weight, $F$ is the fermion number,   $x_j$ and $F_j$ are the fugacity and generators for   flavor symmetries.

The Schur index of $D_p(SU(N))$ takes a very simple form \cite{Song:2017oew}:
\be\label{IDpSUNind}
\cI_{D_p(SU(N))}(q, \bm x)=\PE\Big[\frac{q-q^p} {(1-q)(1-q^p)}\chi_{\text{adj}}^{SU(N)}(\bm x)\Big]~,
\ee
where $\bm x$ is the fugacity for $SU(N)$ flavor symmetry, and $\chi_{\text{adj}}^{SU(N)} $ is the character of $SU(N)$ in the adjoint representation. 
 
   The index of gauged theory can be obtained by taking the product of  the individual matter components  and vector multiplet contributions, and then projecting to gauge invariant sector. For our theory $\cT_{(p,N)}$, the index is given by 
  \be
\cI_{ \cT(p,N)} (q)=\oint [d\bm x ] \; \prod_{j=1}^\ell \cI_{D_{p_j}(SU(N))}(q,\bm x)   \times \cI_{\text{vec}}(q,\bm x)~,
 \ee
where the Haar measure for $SU(N)$ is
\be
[d \bm x ] = \prod_{i=1}^{N-1}\frac{d x_i}{2\pi ix_i} \prod_{j,k=1\atop j\neq k}^N (1-\frac{x_j}{x_k})~,
\qquad x_1x_2\cdots x_N=1~,
\ee
and the vector multiplet  contribution is
\be
\cI_{\vm}(q, \bm x)=\PE\Big[-\frac{2q} {1-q}\chi_{\adj}^{SU(N)}(\bm x)\Big]~, 
\ee
The notation $\PE$ stands for plethystic exponential defined by
\be
\PE[f(x_1, x_2,\cdots) ]=\exp\Big( \sum_{k=1}^\infty \frac{f(x_1^k, x_2^k, \cdots)}{k}\Big)~.
\ee

After combining all the contributions,  the Schur index of     $\cT_{(p,N)}$ AD theory  then takes the following explicit form
\beqn
    \cI_{\cT_{(p,N)}} (q ) &=&
  \oint [ d\bm x]\, \PE \left[\left( -\frac{2q}{1 - q} +\sum_i\frac{q-q^{p_i}} {(1-q)(1-q^{p_i})} \right) \chi_{ \text{adj}}^{SU(N)}(\bm x) \right]~.
\eeqn

On the other hand, the Schur index  of $\cN=4 $ SYM, which can be thought as an $\cN=2$ theory with one vector and one hyper transforming in the adjoint representation,  is:
\beqn\label{SYMIndex}
    \cI_{\mathcal{N}=4\; SU(N)\; \SYM} (q, y)& =& \oint [ d\bm x]\,\cI_{\text{hyper}}(q, y,\bm x)\cI_{\vm}(q, \bm x)
\\& =&
\oint [ d\bm x]\, \PE \left[\left( -\frac{2q}{1 - q} + \frac{q^{\frac{1}{2}}}{1 - q}(y + y^{-1}) \right) \chi_{ \text{adj}}^{SU(N)}(\bm x) \right] ~,
\eeqn
 where the hypermultiplet  contribution is
 \be 
\cI_{\text{hyper}}(q, y,\bm x)=\PE\Big[\frac{q^{\frac12}} {1-q}(y+\frac 1 y)\chi_{\adj}^{SU(N)}(\bm x)\Big]~,
\ee
and $y$ is the fugacity of the $SU(2)_F$ symmetry arising from the $SU(4)_R$ symmetry  of $\cN=4$ SYM.

It is not obvious, but   easy to verify that the   indices of $\cT_{(p,N)}$ and    $\cN=4 $ SYM theories are related in a simple  way
\be\label{TpNIndexrelation}
\cI_{\cT_{(p,N)}}(q)=    \cI_{\mathcal{N}=4\; SU(N)\; \SYM}\big(q^p, q^{\frac p 2-1}\big)
=  \oint [ d\bm x]\, \PE \left[\left( \frac{q+q^{p-1}-2 q^p }{1-q^p}\right) \chi_{ \text{adj}}^{SU(N)}(\bm x) \right] ~.
\ee 
This index relation between AD theories and $\cN=4$ SYM theories was first observed in \cite{Buican:2020moo}. In the special case of $p=3,N=2$, it has   been  understood as a consequence of  the operator map  in VOAs of two theories.

We will use this index relation to derive a simple formula for the Schur index of $\cT_{(p,N)}$ theory. 
For this purpose, we need to first review the known closed formula for the Schur index of $\cN=4$ SYM theory, which turns out to be simpler for $U(N)$ gauge group. 

In \cite{Bourdier:2015wda},  a closed form expression for the unflavored Schur index    of $\cN=4$ $ U(N)$ SYM theory was given
\be
    \cI_{\mathcal{N}=4\;  U(N)\; \SYM} (q )=\frac{1}{\theta(q)} \sum_{n=0}^\infty
    (-1)^n \Bigg[ \PBK{N+n\\ N}+\PBK{N+n-1\\ N}\Bigg] q^{\frac{ n^2+Nn}{2}}~,
\ee
where 
\be
\theta(q) =\sum_{n=-\infty}^\infty (-1)^n q^{\frac{n^2}{2}}=(q;q)_\infty (q^{\frac12};q)_\infty^2~,
\ee
and the $q$-Pochhammer symbol
\be
(x;q)_\infty =\prod_{j=0}^\infty (1-x q^j)~.
\ee

One useful property that we will use later is that in the limit $x\to 1$,  
\be\label{qPochlim}
\lim_{x\to 1}\frac{ (x;q)_\infty }{1-x}=\prod_{j=1}^\infty (1-  q^j)=(q,q)_\infty~.
\ee

 In \cite{Hatsuda:2022xdv}, the generalization to the case of Schur index with flavor fugacity turned on was achieved using the fermi gas approach. For convenience, we introduce $\xi$ via  $y=\sqrt{q}/\xi$, then the  flavored Schur index of $ \mathcal{N}=4$  $U(N)$ SYM theory can be written as
 \be\label{N4index}
  \cI_{\mathcal{N}=4\;  U(N)\; \SYM} \big(q, y=\sqrt{q}/\xi\big) = \Lambda_{U(N)}(q,u,\xi) Z_N(q,u,\xi)~, \qquad
 \ee
 where
 \be\label{LamdUN}
   \Lambda_{U(N)}(q,u,\xi) =   (-1)^N \xi ^{-\frac{1}{2} (N-1) N}  \frac{ \left(\frac{q}{u};q\right)_{\infty } (u;q)_{\infty }}{\left(\frac{q }{u} \xi ^{-N};q\right){}_{\infty } \left(u \xi ^N;q\right){}_{\infty }}~,
 \ee 
 and $Z_N$ is defined via the generating function
 
 \be \label{XiUN}
 \Xi(\mu,q,u,\xi)=
 \sum_{N=0}^\infty  \mu^N Z_N(q,u,\xi)=\prod_{j=-\infty}^\infty \Big(1- \frac{\mu  \xi ^j}{1-u q^j }  \Big)~.
 \ee
 Here $u$ is an arbitrary parameter whose contribution to index cancels out finally. We will specialize  $u$  to different values in order to simplify the formula. 
 
Let us first consider the  limit $u\to 1$, but keep $q, \xi$ generic. Using \eqref{qPochlim}, we find \eqref{LamdUN} simplifies
 \be\label{LmdtN}
    \Lambda_{U(N)}(q,u,\xi) = (1-u)\tilde     \Lambda_{U(N)}(q,\xi) +O((1-u)^2)~, 
    \ee
    where
    \be\label{LmdtNss}
    \tilde   \Lambda_{U(N)}(q, \xi) =   (-1)^N \xi ^{-\frac{1}{2} (N-1) N}  \frac{  (q;q)_{\infty }^2}{\left( q\xi ^{-N};q\right){}_{\infty } \left(  \xi ^N;q\right){}_{\infty }}~.
 \ee
 
  Meanwhile, \eqref{XiUN} becomes
  \be 
 \Xi(\mu,q,u,\xi)=
 \sum_{N=0}^\infty  \mu^N Z_N(q,u,\xi)=
 \Big(1- \frac{\mu   }{1-u  }  \Big)
 \prod_{j=-\infty\atop j\neq 0}^\infty \Big(1- \frac{\mu  \xi ^j}{1-u q^j }  \Big)
 = \frac{1   }{1-u  }  \tilde \Xi(\mu,q ,\xi)+O((1-u)^0)~,
 \ee
 where
   \be \label{XitNss}
\tilde \Xi(\mu,q, \xi)=
 \sum_{N=0}^\infty  \mu^N \tilde Z_N(q, \xi)=
-\mu \prod_{j=-\infty\atop j\neq 0}^\infty \Big(1- \frac{\mu  \xi ^j}{1-  q^j }  \Big)~,
 \ee
 and we have
 \be\label{ZtN}
 Z_N(q,u,\xi)=\frac{1}{1-u} \tilde Z_N(q, \xi)+O((1-u)^0)~,
 \ee
 
 Combining \eqref{LmdtN} and \eqref{ZtN},  the $\cN=4$ $U(N))$ SYM index \eqref{N4index} can be written as
  \be
  \cI_{\mathcal{N}=4\;  U(N)\; \SYM} \Big(q, y=\sqrt{q}/\xi\Big)=\lim_{u\to 1} \Lambda_{U(N)}(q,u,\xi) Z_N(q,u,\xi) =  \tilde Z_{N }(q, \xi) \tilde \Lambda_{U(N)}(q, \xi) ~, \qquad
 \ee
Taking $N=1$, we get
   \be
  \cI_{\mathcal{N}=4\;  U(1)\; \SYM} (q, y=\sqrt{q}/\xi) =\tilde Z_{1 }(q, \xi) \tilde \Lambda_{U(1)}(q, \xi)
     = \frac{  (q;q)_{\infty }^2}{\left( q/\xi  ;q\right){}_{\infty } \left(  \xi ;q\right){}_{\infty }}~,
      \ee
 where we used \eqref{LmdtNss} and \eqref{XitNss}.

With  this $U(1)$ index, we can get the Schur index of $\cN=4$ $SU(N)$ SYM theory:
 \be
  \cI_{\mathcal{N}=4\;  SU(N)\; \SYM} \Big(q, y=\sqrt{q}/\xi\Big)
  =\frac{  \cI_{\mathcal{N}=4\;  U(N)\; \SYM} \Big(q, y=\sqrt{q}/\xi\Big)}{  \cI_{\mathcal{N}=4\;  U(1)\; \SYM} \Big(q, y=\sqrt{q}/\xi\Big)}
     =  \Lambda_{SU(N)}(q,u,\xi) Z_N(q,u,\xi)~, \qquad
 \ee
 where
 \beqn
   \Lambda_{SU(N)}(q,u,\xi) &=&  \frac{  \Lambda_{ U(N)}(q,u,\xi) }{     \cI_{\mathcal{N}=4\;  U(1)\; \SYM} \Big(q, y=\sqrt{q}/\xi\Big) } 
   \\&=&
      (-1)^N \xi ^{-\frac{1}{2} (N-1) N}  \frac{ \left(\frac{q}{u};q\right)_{\infty } (u;q)_{\infty }}{\left(\frac{q }{u} \xi ^{-N};q\right){}_{\infty } \left(u \xi ^N;q\right){}_{\infty }}
    \frac {\left( q/\xi  ;q\right){}_{\infty } \left(  \xi ;q\right){}_{\infty }}{  (q;q)_{\infty }^2}~.
 \eeqn
 
Now we can use the index relation \eqref{TpNIndexrelation} to get the Schur index of $\cT_{(p,N)}$ theory. 
 To distinguish, we will use $Q$ in place of $q$ as the argument for the Schur index of $\cT_{(p,N)}$. Then \eqref{TpNIndexrelation}  implies  
 \be
   \cI_{\cT_{(p,N)}}  (Q) =  \cI_{\mathcal{N}=4\;  SU(N)\; \SYM} (q=Q^p, y=\sqrt{q}/\xi=Q^{p/2-1}) 
    =  \Lambda_{SU(N)}(Q^p,u,Q) Z_N(Q^p,u,Q)~.
 \ee
Therefore we must set   $\xi=Q$ and
 \beqn
   \Lambda_{SU(N)}(Q^p,u,Q) &=&  
      (-1)^N Q ^{-\frac{1}{2} (N-1) N}  \frac{ \left(\frac{Q^p}{u};Q^p\right)_{\infty } (u;Q^p)_{\infty }}{\left(\frac{Q^ {p-N} }{u}  ;Q^p\right){}_{\infty } \left(u Q ^N;Q^p\right){}_{\infty }}
    \frac {\left( Q^{p-1}  ;Q^p\right){}_{\infty } \left(  Q ;Q^p\right){}_{\infty }}{  (Q^p;Q^p)_{\infty }^2}~, \qquad
\\
 \Xi(\mu,Q^p,u,Q) &=&  
 \sum_{N=0}^\infty  \mu^N Z_N(Q^p,u,Q)=
 \Big(1- \frac{\mu/Q   }{1-u/Q^p  }  \Big)
 \prod_{j=-\infty\atop j\neq -1}^\infty \Big(1- \frac{\mu  Q ^j}{1-u Q^{jp} }  \Big)~.
 \eeqn
 
 To proceed further, we take an alternative limit $u\to Q^p$. Using \eqref{qPochlim}, we get
\be \label{SULmd}
  \Lambda_{SU(N)}(\mu,Q^p,u,Q) = (1-Q^p/u)
  \frac{(-1)^N Q^{-\frac{1}{2} (N-1) N} \left(Q;Q^p\right){}_{\infty } \left(Q^{p-1};Q^p\right){}_{\infty }}{  \left(Q^{-N};Q^p\right){}_{\infty } \left(Q^{N+p};Q^p\right){}_{\infty }}
  +O\Big((1-Q^p/u)^2\Big)~,
\ee
  and
  \beqn
   \Xi (\mu,Q^p,u,Q)  &=&  
 \sum_{N=0}^\infty  \mu^N Z_N(Q^p,u,Q)=
 - \frac{\mu/Q   }{1-u/Q^p  }    \Omega(\mu,Q^p, Q)   +O\Big((1-Q^p/u)^0\Big)~, \qquad
 \eeqn
  where
\be
 \Omega(\mu,Q^p, Q)= \sum_{N=0}^\infty  \mu^{ N-1} J_N(Q^p, Q ) = 
 \prod_{j=-\infty\atop j\neq -1}^\infty \Big(1- \frac{\mu  Q ^j}{1-  Q^{(j+1)p} }  \Big)
 = \prod_{j=-\infty\atop j\neq 0}^\infty \Big(1- \frac{\mu  Q ^{ j-1}}{1-  Q^{jp} }  \Big)~.
 \ee
This implies
  \be\label{ZNSUN}
  Z_N(Q^p,u,Q)=
  - \frac{1/Q   }{1-u/Q^p  }  J_N(\mu,Q^p, Q)   +O\Big((1-Q^p/u)^0\Big)~.
  \ee
   Combining \eqref{SULmd} and \eqref{ZNSUN} together, we get
 \be\label{cTpNs}
    \cI_{\cT_{(p,N)}}  (Q) =  \frac{(-1)^N Q^{-\frac{1}{2} (N-1) N}   \left(Q;Q^p\right){}_{\infty } \left(Q^{p-1};Q^p\right){}_{\infty }}{  \left(Q^{-N};Q^p\right){}_{\infty } \left(Q^{N+p};Q^p\right){}_{\infty }} J_N/Q~.
 \ee
 To further simplify the expression, we  introduce $\lambda=-\mu/Q$ and $I_N$ via
 \be
\Omega= \sum_{N=0}^\infty  \lambda  ^{ N-1} I_N (Q)= \sum_{N=0}^\infty  (-\lambda Q)^{ N-1} J_N 
 = \prod_{j=-\infty\atop j\neq 0}^\infty \Big(1+ \frac{\lambda  Q ^{ j }}{1-  Q^{jp} }  \Big)~.
 \ee
 It is easy to see that $  I_N= (-  Q)^{ N-1} J_N  $. Substituting it to \eqref{cTpNs}, we get
 
  \beqn
    \cI_{\cT_{(p,N)}}  (Q) &=&  \frac{(-1)^N Q^{-\frac{1}{2} (N-1) N}   \left(Q;Q^p\right){}_{\infty } \left(Q^{p-1};Q^p\right){}_{\infty }}{  \left(Q^{-N};Q^p\right){}_{\infty } \left(Q^{N+p};Q^p\right){}_{\infty }} I_N/(- Q)^{ N-1}/Q
\\&=&
 - \frac{ Q^{-\frac{1}{2} (N+1) N}   \left(Q;Q^p\right){}_{\infty } \left(Q^{p-1};Q^p\right){}_{\infty }}{  \left(Q^{-N};Q^p\right){}_{\infty } \left(Q^{N+p};Q^p\right){}_{\infty }} I_N~.  
 \eeqn
 
For convenience of notation, we change variables back to   $q$ and $\mu$ and get
 \beqn
  \cI_{\cT_{(p,N)}}(q)&=& C(q)
 Z_{\cT_{(p,N)}}(q)~, \qquad  
\eeqn
where
\beqn
   C(q)= -\frac{q^{-\frac{1}{2} N (N+1)} \left(q;q^p\right){}_{\infty } \left(q^{p-1};q^p\right){}_{\infty }}{\left(q^{-N};q^p\right){}_{\infty } \left(q^{N+p};q^p\right){}_{\infty }}~,
   \qquad
 \sum_{N=1}^\infty \mu^{N-1}    Z_{\cT_{(p,N)}}(q)&=&\prod_{j=-\infty\atop j\neq 0}^\infty
  \Big( 1+ \frac{\mu  q^j}{1-q^{j p}}  \Big)~.\qquad
 \eeqn
 Surprisingly, the formula can be further simplified for the    $\cT_{(p,N)}$ AD   theories studied here. In particular, the values of $p,N$ are restricted such that $p=2,3,4,6$ and $\gcd(p,N)=1$. So in all cases we always    have $N=rp\pm 1$ for $r\in \bZ$. If $N=rp+1$, we have
 \beqn
 \frac{  \left(q;q^p\right){}_{\infty } \left(q^{p-1};q^p\right){}_{\infty }}{\left(q^{-N};q^p\right){}_{\infty } \left(q^{N+p};q^p\right){}_{\infty }}
 &= &
 \frac{  \left(q;q^p\right){}_{\infty } \left(q^{p-1};q^p\right){}_{\infty }}{  \left(q^{(r+1)p+1};q^p\right){}_{\infty } \left(q^{-rp-1};q^p\right){}_{\infty }}
\\ &= &
\prod_{j=0}^r\frac{1-q^{jp+1}}{1-q^{-jp-1}}
=\prod_{j=0}^r (-q^{jp+1} )
 \\ &= &
 (-1)^{r+1}q^{\sum_{j=0}^r jp+1} 
  \\ &= &
   (-1)^{r+1}q^{ (rp+2)(r+1)/2}~,
 \eeqn
therefore  
 \be
C(q)=  -q^{-\frac{1}{2} N (N+1)}  (-1)^{r+1}q^{ (rp+2)(r+1)/2}
=
 (-1)^r q^{-\frac{1}{2} (p-1) r (p r+2)}
 = (-1)^rq^{-\frac{1}{2} \left(N^2-1\right) \left(1-\frac{1}{p}\right)}~.
 \ee
 Similarly, if $N=rp-1$, we find
 \be
C(q)= - (-1)^r q^{-\frac{1}{2} (p-1) r (p r-2)}= -(-1)^rq^{-\frac{1}{2} \left(N^2-1\right) \left(1-\frac{1}{p}\right)}~.
\ee

In both cases, we have
 \be
 C(q)=
(-1)^r s q^{-\frac{1}{2} \left(N^2-1\right) \left(1-\frac{1}{p}\right)}=(-1)^r s q^{-\frac{1}{2} c^{\cT_{(p,N)}}}
=(-1)^{\floor{\frac{N}{p}}} q^{-\frac{1}{2} c^{\cT_{(p,N)}}} ~,
 \qquad N=rp+s~, \quad s=\pm 1~.
 \ee
 where we used \eqref{TpNcentralcharge}. 
 
 We can then redefine $\cZ_{ \cT_{(p,N)}}(q) =(-1)^{\floor{\frac{N}{p}}} Z_{ \cT_{(p,N)}}(q) $ and get the final formula for the Schur index
  \be\label{TpNSchur}\boxed{ 
  \cI_{\cT_{(p,N)}}(q)=  
 q^{-\frac{1}{2} c^{\cT_{(p,N)}}} \cZ_{ \cT_{(p,N)}}(q) 
= 
 q^{ \frac{1}{24} c_{2d}^{\cT_{(p,N)}}} \cZ_{ \cT_{(p,N)}}(q) ~,
 }
\ee 
where $\cZ_{ \cT_{(p,N)}}(q)$ is given by the generating function
\be \label{TpNZNN}\boxed{ 
 \sum_{N=1}^\infty \mu^{N-1} (-1)^{\floor{\frac{N}{p}}}  \cZ_{\cT_ {(p,N)}}(q)
 =
  \prod_{j=-\infty\atop j\neq 0}^\infty
   \Big( 1+ \frac{\mu  q^j}{1-q^{j p}}  \Big)
  ~.} 
 \ee 
 Equivalently, we can write the generating function as
 \be \label{TpNZNN23} 
 \sum_{N=1}^\infty \mu^{N-1} (-1)^{\floor{\frac{N}{p}}}  \cZ_{\cT_ {(p,N)}}(q)
  =\prod_{j=1}^\infty \Big( 1+\frac{\mu  q^j \left(1-q^{j (p-2)}\right)}{1-q^{j p}}-\frac{\mu ^2 q^{j p}}{\left(1-q^{j p} \right)^2} \Big) ~. 
 \ee 
 
 Obviously, \eqref{TpNSchur} has exactly the factorized form in  \eqref{SchurPS}. The factorization emerges naturally from our derivation.~\footnote{One may complain that  $(-1)^{\floor{\frac{N}{p}}}$ is not natural. But this is mild as it is just a sign. Furthermore, this sign is actually important  for  $ \cZ_{\cT_ {(p,N)}}(q)$ in \eqref{TpNZNN} to have positive coefficients. This point is particularly obvious for $p=2$, see \eqref{T2N}. }
 As we discussed in the introduction, the Schur partition function $\cZ_{ \cT_{(p,N)}}(q)$ is identified with the vacuum character of the corresponding chiral algebra, and is expected to have nice modular properties. 
 The $\cZ_{ \cT_{(p,N)}}(q)$ defined via \eqref{TpNZNN} is very simple,   suggesting that it defines an elementary  function with nice modular properties and deserves further studies.
   
The Schur partition function     $\cZ_{ \cT_{(p,N)}}(q)$ can be more explicitly written as
 \be
\cZ_{ \cT_{(p,N)}}(q)=  (-1)^{\floor{\frac{N}{p}}} 
 \sum_{m_1<m_2 <\cdots <m_{N-1}\atop  m_i\in \bZ, m_i\neq 0}\prod_{j=1}^{N-1}\frac{   q^{m_j}}{1-q^{m_j p}}~.
  \ee
  
   When $p=2$, it is easy to see that $m_i$  and $m_j=-m_i$  must appear in pair in each summand, otherwise,  the two summands with  $m_i$ and $ -m_i$ would cancel. One can also see this point from the   expression in \eqref{TpNZNN23}.  Therefore, we should have $N=2k+1$, and
    \be\label{T2N}
\cZ_{ \cT_{(2,2k+1)}}(q)=  (-1)^k 
 \sum_{0<m_1<m_2 <\cdots <m_{k}<\infty}\prod_{j=1}^{k}   \frac{- q^{2m_j}}{(1-q^{2m_j})^2}
 = \sum_{0<m_1<m_2 <\cdots <m_{k}<\infty}\prod_{j=1}^{k}   \frac{  q^{2m_j}}{(1-q^{2m_j})^2}~.
  \ee
  Note that the sign factor $(-1)^{\floor{\frac{N}{p}}} =(-1)^k$ precisely cancels the sign from the product, rendering a function with positive coefficients in $q$.
 In this case, $\cZ_{ \cT_{(2,2k+1)}}(q)$ turns out to be related to the known function via $\cZ_{ \cT_{(2,2k+1)}}(q) =A_k(q^2)$, \footnote{This special case of $p=2$ was noticed before in \cite{Kang:2021lic}.} where 
  
 \be\label{Akq}
A_k(q)
= \sum_{ 0<m_1<m_2\cdots <m_k<\infty}    
  \frac{ q^{ m_1+\cdots  m_k}}{ (1- q^{  m_1}  )^2\cdots (1- q^{ m_k}  )^2}
 \ee
 is known as the MacMahon’s generalized ‘sum-of-divisor’ function \cite{https://doi.org/10.1112/plms/s2-19.1.75}. It satisfies the  recursion relation \cite{andrews2010macmahons}
 \be\label{Akrecursion}
 A_k(q) =\frac{1}{2k(2k+1)}\Big[\big(6 A_1(q) +k(k-1) \big) A_{k-1}(q) -2q \frac{d}{dq} A_{k-1}(q) \Big]~,
 \ee
 and
 \be\label{Akq1}
 A_1(q)= \frac12 \mathbb E_{2}(q)+\frac{1}{24 }~.
 \ee
 Here $\bE_{2k}$ are Eisenstein series \eqref{Einseries}. 
This implies  $A_k(q)$ is a quasi-modular form, and can be written as the polynomial   of Eisenstein series $\bE_2,\bE_4,\bE_6$. See  appendix \ref{modformMLDE} for   definitions of various concepts.~\footnote{However,   $\cZ_{ \cT_{(2,2k+1)}}(q)$ seems not to be a quasi-modular form. One can numerically check that $\cZ_{ \cT_{(2,2k+1)}}(q)$ can not be written as the polynomial   of $\bE_2,\bE_4,\bE_6$ with finite degree. Nevertheless, $\cZ_{ \cT_{(2,2k+1)}}(q)$  still enjoys nice modular properties based on the physical expectation of SCFT/VOA correspondence, and satisfies MLDE as we will discuss later. }
  
 For $p=3,N=2$, we have
   \beqn\label{SchT32}
\cZ_{ \cT_{(3,2)}}(q)&=&   
 \sum_{m  \in \bZ, m\neq 0} \frac{   q^{m}}{1-q^{3m }}
 =  \sum_{m  \in \bZ_+} \frac{   q^{m }-q^{2m }}{1-q^{3m }}
  = \sum_{m =1}^\infty \frac{   q^{m } }{1+q^{ m }+q^{2m}}
 \\ &=&
q\Big( 1+q^2+q^3+2 q^6+q^8+q^{11}+2 q^{12}+q^{15}+\cdots \Big)~.
  \eeqn
   
    For $p=4,N=3$, we have 
  \beqn\label{T43SchuPT}
\cZ_{ \cT_{(4,3)}}(q) &=&   \sum_{m>n\ge 1}\frac{q^{m+n}+q^{3 (m+n)}}{\left(1-q^{4 m} \right) \left(1-q^{4 n} \right)}
  -\sum_{m,n\ge 1}\frac{q^{3 m+n}}{\left(1-q^{4 m} \right) \left(1-q^{4 n} \right)}
  \\ &=& 
q^3 \Big(1+q^2+q^3+2 q^4+3 q^6+q^7+3 q^8+q^9+3 q^{10}+2 q^{11}+6 q^{12}+4 q^{14}+3 q^{15}+\cdots \Big) ~. \qquad\qquad
  \eeqn
  
  Note that    \eqref{TpNZNN} is derived from physics and    the values of $p$ and $N$ are subject to various conditions.  But at the level of mathematics, \eqref{TpNZNN}  applies to all $p$ and $N$. Some may be trivial. For example, one can   check that
   \be
\cZ_{ \cT_{(3,3)}}(q)=   \sum_{m>n\ge 1}\frac{q^{m+n}+q^{2 (m+n)}}{\left(1-q^{3 m} \right) \left(1-q^{3 n} \right)}
  -\sum_{m,n\ge 1}\frac{q^{2 m+n}}{\left(1-q^{3 m} \right) \left(1-q^{3 n} \right)}=0~.
  \ee
  and more generally we expect   $\cZ_{\cT_{(p,jp)}}=0$ for $j\in \mathbb N$. On the other hand for $p=4,N=2$,
$\cZ_{ \cT_{(4,2)}}(q)$ is not zero and furthermore satisfies the MLDE of weight 18.   It would be interesting to study these properties further.

   \section{Modular linear differential equation  
   } \label{TpNMLDE}
 In this section,   we will study the modular linear differential equations in  $\cT_{(p,N)}$ AD theories.  In the simplest case of $\cT_{(3,2)}$ whose VOA is known, we will derive the corresponding  MLDE using the formalism in Appendix~\ref{modformMLDE} and \ref{Zhurecursion}. In other cases, we will present the modular linear differential equations satisfied by Schur partition function, which are found through numerics.  We will also discuss the solutions to the MLDEs and  their modular transformation properties.
   
 \subsection{ $\cT_{ (3,2)}$} \label{T32MLDE}
   
  The chiral algebra of $\cT_{(3,2)}$ AD theory  is discussed     in \cite{Buican:2020moo}   and given by the $\mathcal A(6)$ algebra   \cite{feigin2007fermionic,feigin2008characters}.
It contains 3 generators, denoted by $T, \Phi, \widetilde\Phi$, whose conformal dimensions are 2, 4, 4, respectively. While the first generator $T$ is the stress tensor, the latter two $\Phi, \widetilde\Phi$ are fermionic Virasoro primary operators.  The OPEs among them are given by
\beqn
T(z)T (0)  &\sim& \frac{-12}{z^4}+\frac{2T}{z^2}+\frac{T'}{z} ~,\\
T(z)\Psi (0)  &\sim&  \frac{4\Psi}{z^2}+\frac{\Psi'}{z} ~, \\
T(z)\widetilde\Psi (0)  &\sim& \frac{4\widetilde\Psi}{z^2}+\frac{\widetilde\Psi'}{z}  ~,\\
\Psi(z)\widetilde\Psi (0)  &\sim&
-\frac{6 }{z^8} +\frac{2 T}{z^6} + \frac{T'}{z^5} +\frac{3 \left(T''-T^2\right)}{7 z^4}+\frac{2 T^{(3)}-9 T' T}{21 z^3}
\nonumber\\&&
+\frac{-48 \left(T'\right)^2-84 T ''T+36 T^3+7 T^{(4)}}{420 z^2}+\frac{60 \left(-5 T'' T'+6T' T^2  -2 T^{(3)} T\right)+7 T^{(5)}}{2800 z} ~, \qquad\qquad
\eeqn
where we ignore the coordinate of operators on the RHS, which is 0.
Actually the full OPE can be easily bootstrapped using the associativity of OPEs and the information about the conformal dimensions of these operators.  
 
The mode expansion of these operators are
 \be 
 T(z)   
  =\sum_{n\in \bZ}L_{n}z^{-2-n}~,\qquad
   \Phi(z)   
  =\sum_{n\in \bZ}\Phi_{n}z^{-4-n}~,\qquad
   \widetilde\Phi(z)   
  =\sum_{n\in \bZ}\widetilde\Phi_{n}z^{-4-n}~.
 \ee
Using the previous OPEs, we manage  to derive the commutation relations of these modes 
 %
\beqn
[L_m,L_n]&=& (m-n)L_{m+n}-2(m^3-m)\delta_{m+n,0}~,
\\ 
{} [L_m,\Phi_n]&=& (3m-n)\Phi_{m+n} ~,
\\
{} [L_m,\widetilde\Phi_n]&=& (3m-n)\widetilde\Phi_{m+n} ~,
\\
\{\Psi_m, \widetilde\Psi_n\}&=&
\frac{1}{840} n \left(n^2 \left(n^2-7\right)^2-36\right) \delta _{0,m+n}-\frac{1}{84} (m-n) \left(m^2-m n+n^2-7\right) \Lambda_ { m+n}
\nonumber\\&&
+\frac{1}{1680}(m-n) \Big(3 (m+n)^4-14 m n \left(m^2+m n+n^2\right) -39 (m+n)^2+98 m n+108 \Big) L_{ m+n} 
\nonumber\\&&
-\frac{5}{112} (m-n) \widetilde\Lambda_ { m+n}+\frac{7}{80} (m-n) \Upsilon_{m+n}~,
\label{Phicomm}
\eeqn
 where  \footnote{The $\Lambda,\widetilde\Lambda,\Upsilon$ constructed in this way have   simple commutation relations with stress tensor. E.g. $ [L_m,\Lambda_n]=\frac{1}{30} (5 c+22) m \left(m^2-1\right)L_{m+n} +(3m-n) \Lambda_{m+n}$.
}
 \beqn\label{lmda2}
\Lambda&=&T^2-\frac{3}{10}T''~,
\\  
\widetilde\Lambda&=&T^3 -\frac{1}{3} (T')^2 -\frac{19}{30}T'' T-\frac{1}{36}T''''~, \qquad
\Upsilon=T^3 -\frac17 (T')^2 -\frac{11}{14}T'' T-\frac{19}{588} T''''~.
\eeqn

 With the explicit OPEs, we can try to find the null operators. See appendix \ref{MLDET23} for detailed discussions.  In particular, we find a null operator at dimension 10:
\beqn
 N_{10}&=&
T^5+\frac{4}{3}   T^{(3)} T' T-\frac{10}{3}   \left(T'\right)^2 T^2 -\frac{10    }{3} T''T^3+2  T'' \left(T'\right)^2 +\frac{9}{4}  \left(T''\right)^2T
 -\frac{1}{9} \left(T^{(3)}\right)^2
\nonumber \\&&
 -\frac{1}{9}   T^{(4)} T^2 -\frac{1}{12}  T^{(4)} T''-\frac{1}{30}  T^{(5)}  T'-\frac{13}{360}   T^{(6)}T+\frac{ 1}{5040}T^{(8)}
+\frac{70  }{9}  \Psi ''\tilde{\Psi } +\frac{140 }{3}   \tilde{\Psi }''  \Psi~.
\qquad\quad
\eeqn
 
 In terms of modes, the corresponding null state   is
  \beqn
\cN_{10} &=&
 \Big( L_{-2}^5-\frac{20}{3} L_{-4} L_{-2}^3-\frac{10}{3} L_{-3}^2 L_{-2}^2-\frac{8}{3} L_{-6} L_{-2}^2+9 L_{-4}^2 L_{-2}-26 L_{-8} L_{-2}+8 L_{-5} L_{-3} L_{-2}
\nonumber \\ &&\;\;
 -4 L_{-5}^2+4 L_{-4} L_{-3}^2+8 L_{-10}-4 L_{-6} L_{-4}-4 L_{-7} L_{-3}+\frac{280}{3}  \tilde  \Psi   _{-6}  \Psi   _{-4}+\frac{140}{9} \Psi   _{-6} \tilde \Psi   _{-4} 
 \Big)\Omega ~.
 \qquad \qquad
  \eeqn
This null state  has the form of \eqref{nullT}, namely $(L_{-2})^5 \Omega  \in C_2(\cV)$ \eqref{C2V}. The presence of such a kind of null operator enables us to derive the MLDE using Zhu's recursion relation \eqref{Zhu}\eqref{Zhu2}. We defer the detailed discussions and derivations to the appendix \ref{MLDET23}, and only provide the final result here. The resulting MLDE we find takes the following simple form       

  \be\label{MDeqT322}
 \Big[ D_q^{(5)} -140 \mathbb E_{4}   D_q^{(3)} -700 \mathbb E_{6}   D_q^{(2)}-2000   \mathbb E_{4}^2   D_q^{(1)} \Big]
   \cZ_{\cT_{ (3,2)}}=   0~,
    \ee
 where    $\bE_{2k}$ are Eisenstein series \eqref{Einseries},  and $D_q^{(k)}$ are modular covariant differential operators \eqref{modcovder}. See appendix \ref{modformMLDE} for   review and discussion on the notations and properties. 
 
 The explicit and simple expression of $   \cZ_{\cT_{ (3,2)}}$ is given in \eqref{SchT32}. One can then numerically verify that the above MLDE is indeed true. 
 
 We would like to find the full set of solutions to MLDE, which correspond to the characters of some modules in the corresponding $\mathcal A(6)$ chiral algebra. We can use the following ansatz
 \be\label{ansatz}
 \chi_b =q^b (1+a_1 q+a_2 q^2 +\cdots)~, \qquad b=-\frac{c_{2d}}{24}+h~.
 \ee
 Then we get the following indicial equation
 \be
 (1-3 b)^2 (b-1) b^2=0~, \qquad \to \qquad b= 0,\; 0,\; \frac13,\; \frac13,\; 1~.
 \ee
 
Due to the degeneracy and integral spacing of the roots, the  ansatz \eqref{ansatz} is not valid in general. instead  we should   use the following ansatz: 
 \be
 \chi_{b}=q^{b}\sum_{ j=0}^{N_b}\sum_{i=0}^\infty a_{ij}(\log q)^jq^i ~,
  \ee
  where $N_b$ depends on the structure of roots to the indicial equation.   See appendix~\ref{MLDEsol} for more discussions.

 We then find the following  set of solution to MLDE \eqref{MDeqT322}
 \beqn 
b=0:\qquad \chi_0 &=& 1~,
\\
b=0:\quad \;\;  \chi_0^{\log}&=&\log q\Big( 1+6 q+6 q^3+6 q^4+12 q^7+6 q^9+6 q^{12}+12 q^{13}+6 q^{16}+12 q^{19}
 +\cdots 
  \Big)~,\qquad\qquad 
\\
 b=\frac13: \qquad \chi_{\frac13}&=&
 q^{\frac13} \left(1+q+2 q^2+2 q^4+q^5+2 q^6+q^8+2 q^9+2 q^{10}+2 q^{12}+2 q^{14}+3 q^{16}
  +\cdots 
 \right)~,
\\
  b=\frac13: \quad \;\;   \chi_{\frac13}^{\log}&=& \chi_{\frac13} \log q~,
\\
 b=1: \quad \;\; \; \;
 \chi_1 &=&
 \cZ_{\cT_{(3,2)}}=q+q^3+q^4+2 q^7+q^9+q^{12}+2 q^{13}+q^{16}+2 q^{19}+2 q^{21}+q^{25}
  +\cdots~. 
 \eeqn

 Note that since there is no constant term in MLDE \eqref{MDeqT322}, namely the coefficient of $ D_q^{(0)} $ is zero,   $\chi_0=1$ is  also a solution. 
One can also see that
 \be\label{chi0p}
   \chi_0^{\log}(\tau)=\log q( 1+6 \chi_1)=2\pi i \tau \Big( 1+6 \chi_1(\tau) \Big)~.
 \ee

The five linearly independent solutions above are the full solutions to the MLDE. They are also the characters of some modules in the corresponding chiral algebra. And we expect that they form a  (weakly holomorphic logarithmic) vector-valued modular  under modular transformation. More precisely, we can consider the vector of solutions 
 \be
 \bm\chi=\Big(\chi_0,  \quad   \chi_0^{\log}, \quad \chi_1,\quad \chi_{\frac13}, \quad \chi_{\frac13}^{\log}  \Big)^T~.
 \ee
 Then, under modular $S$- and $T$-transformation, we expect 
 \be\label{STtsf}
  \bm\chi(-\frac 1\tau)=S \bm \chi(\tau)~, \qquad
    \bm\chi( \tau+1)=T \bm \chi(\tau)~, \qquad
 \ee
 where $S,T$ are modular $S$ and $T$ matrices. 
 
 The   modular $T$ matrix can be easily derived. In the absence of $\log $ term, $\chi_b$ transforms to $e^{2\pi i b}\chi_b$. In the presence of $\log$ term,  $\chi_b$ gets mixed with other items in $\bm \chi$, and  the modular  $T$ matrix is not diagonal anymore. For example, it is easy to see that
  \be
  \chi_0^{ \log  }(\tau+1)=   \chi_0^{\log}(\tau)+2\pi i ( 1+6 \chi_1(\tau) )~, \qquad
    \chi_{\frac13}^{ \log}(\tau+1)= 2\pi ie^{\frac{2 i \pi }{3}}\chi_{\frac13}(\tau)+ e^{\frac{2 i \pi }{3}}\chi_{\frac13}^{\log}(\tau)~.
 \ee
 As a result, we find the modular $T$ matrix
 \be
 T=\left(
\begin{array}{ccccc}
 1 & 0 & 0 & 0 & 0 \\
 2 \pi  i & 1 & 12   \pi  i & 0 & 0 \\
 0 & 0 & 1 & 0 & 0 \\
 0 & 0 & 0 & e^{\frac{2 i \pi }{3}} & 0 \\
 0 & 0 & 0 & 2\pi i e^{\frac{2 i \pi }{3}}    & e^{\frac{2 i \pi }{3}} \\
\end{array}
\right)~.
 \ee
 
  The   modular $S$ matrix  is generally more complicated to compute. Fortunately, we find that some entries in   $\bm \chi$  reduce to known functions whose modular transformation is understood.  
   By numerically  computing the solution  $ \chi_{\frac13}$ to very high order~\footnote{We have verified it up to the order $q^{140}$.}, we find that it actually can be written as  
 \be\label{chi13}
 \chi_{\frac13}(\tau)=q^{\frac13}\frac{\left(q^3;q^3\right)_{\infty }^3}{(q;q)_{\infty }}
  =\frac{\eta(3\tau)^3}{\eta(\tau)}~, \qquad  q\equiv e^{2\pi i}~,
   \ee
 where $ \eta(\tau)=q^{\frac{1}{24}}(q,q)_\infty=q^{\frac{1}{24}}\prod_{j=1}^\infty(1-q^j)$ is the Dedekind eta function. Under modular transformation, the Dedekind eta function transforms as
 \be
 \eta(-1/\tau)=\sqrt{-i\tau}\eta(\tau)~, \qquad 
 \eta(\tau+1) =e^{\frac{\pi i}{12}}\eta(\tau)~.
 \ee
 With the help of these formulae, we can easily find that under $S$-transformation, the solution \eqref{chi13} becomes 
 \be 
 \chi_{\frac13}(-\frac1 \tau)=-\frac{i \tau  }{3 \sqrt{3}  } \frac{ \eta \left(\frac{\tau }{3}\right)^3}{\eta (\tau )}
 =-\frac{\log (q)}{6 \sqrt{3} \pi }\frac{ \eta \left(\frac{\tau }{3}\right)^3}{\eta (\tau )}
 =\frac{1}{2 \sqrt{3} \pi }\chi_{\frac13}^{\log}(\tau)-\frac{1}{6 \sqrt{3} \pi }  \chi_0^{\log}(\tau)~.
 \ee
  where we write the $S$-transformed character as the linear combination of the original untransformed characters in light of \eqref{STtsf}; this can be easily achieved by comparing the coefficients of their $q$-expansions. 
  
 Similarly, $  \chi_{\frac13}^{\log}$ transforms as
 \be
  \chi_{\frac13}^{\log}(-\frac1 \tau)=-\frac{2 \pi   }{3 \sqrt{3}  }\frac{ \eta \left(\frac{\tau }{3}\right)^3}{\eta (\tau )}
  =
 -\frac{4 \pi   }{\sqrt{3}}\chi_1(\tau)+\frac{2 \pi   }{\sqrt{3}}\chi_{\frac13}(\tau)-\frac{2 \pi }{3 \sqrt{3}}\chi_0(\tau)~.
 \ee

  The modular $S$ matrix has to satisfy   the condition    $S^2=1$. Imposing such a constraint, we are able to determine $S$ matrix completely: \footnote{One can rescale the character by replacing $\log q$ with $    \log q/(6\sqrt 3\pi)$ and get the modular $S$  matrix   with rational coefficients. But this would introduce the factor $\sqrt 3$ to  $T$ matrix.
  %
 }
 \be\label{Smat}
 S=\left(
\begin{array}{ccccc}
 1 & 0 & 0 & 0 & 0 \\
 -\frac{2 \pi }{\sqrt{3}} & 0 & -4 \sqrt{3} \pi  & -4 \sqrt{3} \pi  & 0 \\
 -\frac{1}{6} & -\frac{1}{12 \sqrt{3} \pi } & 0 & 0 & -\frac{1}{2 \sqrt{3} \pi } \\
 0 & -\frac{1}{6 \sqrt{3} \pi } & 0 & 0 & \frac{1}{2 \sqrt{3} \pi } \\
 -\frac{2 \pi }{3 \sqrt{3}} & 0 & -\frac{4 \pi }{\sqrt{3}} & \frac{2 \pi }{\sqrt{3}} & 0 \\
\end{array}
\right)~. 
 \ee
 Note that the $S$ matrix is not symmetric.
 It is easy to verify that $S^2=(ST)^3=1$, as required by modular $S$ and $T$ matrices. \footnote{More generally, the condition is $S^2=(ST)^3=C$ where $C$ is the charge conjugation matrix. }
 
 As a result, we find
 \be\label{chi1tau}
 \chi_1(-\frac 1 \tau)= -\frac{1}{6}  \chi_0(\tau)  -\frac{1}{12 \sqrt{3} \pi } \chi^{\log}_0(\tau)    -\frac{1}{2 \sqrt{3} \pi }\chi_ {\frac13}^{\log}(\tau)  ~.
 \ee
 Using \eqref{chi0p}, we then find
  \be
   \chi_0^{\log}(-\frac1\tau )=  \frac{-2\pi i}{  \tau} \Big( 1+6 \chi_1(-\frac1\tau) \Big)
 =\frac{i}{\sqrt 3 \tau}   \chi_0^{\log}(\tau)+\frac{2\sqrt 3 i}{  \tau} \chi_{\frac13}^{\log} (\tau)
 =-\frac{2\pi}{\sqrt 3}\chi_0(\tau)-4\sqrt 3 \chi_1(\tau)-4\sqrt 3 \chi_{\frac13}(\tau)~,
 \ee
which agrees with \eqref{Smat}. This thus provides a strong consistency check of our results.
  
 \subsection{ $\cT_{ (2,3)}$} 
 For the $\cT_{ (2,3)}$ AD theory, the corresponding VOA is not known. So we will resort to numerics to find the corresponding MLDE. The Schur partition function is given by \eqref{T2N}:
 \be
\cZ_ {\cT_{ (2,3)} }=\sum_{m=1}^\infty \frac{q^{2m}}{(1-q^{2m})^2}~.
 \ee
 This simple expression allows one to numerically compute the series expansion to very high order efficiently. 
 Furthermore, the SCFT/VOA correspondence indicates that $\cZ_ {\cT_{ (2,3)} }$ satisfies the MLDE of specific weight, say $2k$:
 \be\label{MLDET232}
   \cD_q^{(k)}\cZ_ {\cT_{ (2,3)} }=0~,
 \ee
 where  $   \cD_q^{(k)}$ is the modular linear differential  operator  and transforms covariantly under modular transformations. More explicitly, it takes the following form \eqref{MLDOs}
 \be
\cD_q^{(k)}=D_q^{(k)}+\sum_{r=1}^k f_r (\tau) D_q^{(k-r)}~, \qquad f_r (\tau) \in \cM_{2r}(\Gamma)~.
 \ee
 Here $\cM_{2r}(\Gamma)$ denotes the modular form of weight $2r$, and is freely generated by $\bE_4$ and $\bE_6$. So $f_r$ can be written as the polynomial of  $\bE_4$ and $\bE_6$ with finitely many undetermined coefficients. Therefore, the MLDO $\cD_q^{(k)}$ is almost fixed completely by modular covariance, up to a finite number of coefficients.  To fix these coefficients, we can consider the  series expansion  of both $\cD_q^{(k)}$ and $\cZ_ {\cT_{ (2,3)} }$ in $q$, and check MLDE in \eqref{MLDET232}. Starting with $k=1$ in  \eqref{MLDET232}, one can check whether it has is a solution. If yes, we then find the desired MLDE, otherwise we increase $k$ and repeat the same procedure. This then offers an efficient   way to find MLDE numerically.

  After implementing the algorithm above, we find that up to the order   $q^{100}$,  there is  indeed a MLDO of weight 24  which annihilates the Schur partition function $\cZ_ {\cT_{ (2,3)} }$, namely  
  \beqn
   \cD_q^{(12)}\cZ_ {\cT_{ (2,3)} }=0~,
  \eeqn
  where
  \beqn
   \cD_q^{(12)}&=&
    D_q^{(12)}-1510 \mathbb{E}_4D_q^{(10)} -55440 \mathbb{E}_6D_q^{(9)}
    -233400 \mathbb{E}_4^2D_q^{(8)}+2364600  \mathbb{E}_4 \mathbb{E}_6 D_q^{(7)}
  \nonumber  \\&& 
    +2000 \left(31228 \mathbb{E}_4^3-41013 \mathbb{E}_6^2\right)D_q^{(6)}
         +1422624000  \mathbb{E}_4^2 \mathbb{E}_6 D_q^{(5)}
  \nonumber       \\&&
        +\left(3925360000 \mathbb{E}_4^4+40438916000 \mathbb{E}_4 \mathbb{E}_6^2\right)D_q^{(4)}
        \nonumber    \\&&
        + \left(420470400000 \mathbb{E}_4^3 \mathbb{E}_6    +344509200000 \mathbb{E}_6^3\right)D_q^{(3)}
\nonumber    \\&&
      +  \left(1168824000000 \mathbb{E}_4^5
        +7510426000000 \mathbb{E}_4^2 \mathbb{E}_6^2\right)D_q^{(2)}
 \nonumber   \\&&
        +\left(23905224000000 \mathbb{E}_4^4 \mathbb{E}_6+31682361200000 \mathbb{E}_4 \mathbb{E}_6^3\right)D_q^{(1)}~.
  \eeqn
  Note that as in the case of $\cT_{(3,2)}$ AD theory, there is also no constant term in $   \cD_q^{(12)}$ here. 
  
  The   MLDE above  is  a differential equation of order 12, so it should have other 11 solutions  corresponding to other modules of the VOA, in addition to the vacuum character  $\cZ_ {\cT_{ (2,3)} }$.
  To find the rest of solutions, we can use the   ansatz
$ \chi_i =q^b (1+a_1 q+\cdots) $ and substitute it into $  \cD_q^{(12)}\chi_i=0$. At leading order in $q$, this gives the  indicial equation  
  \be\label{indT23}
(b-2) (b-1)^3 b^3 (2 b-1)^3 \left(8 b^2-36 b+39\right)=0~.
  \ee
The solutions can be easily found to be (ignoring multiplicity)
  \be
  b=0,\; 1/2,\; 1,\;  2,\; \frac14 (9 - \sqrt 3),\; \frac14 (9 + \sqrt 3)~.
  \ee
  In particular, note that   $b=0$ is the minimal root and has degeneracy 3.
  
  In principle, one can proceed further and find the full set of solutions order by order in $q$. 
  In practice, this computation is tedious as the MLDO is a differential operator of very high order being 12. 
The computation is further complicated by the degeneracies of the roots which means logarithmic term in the solutions.  Given these complications, we will not study the explicit solutions here, and just be content with the indicial equation. As we will discuss later, the indicial equation is already useful enough and can be used to understand the high temperature limit of the Schur index / partition function.

 \subsection{ $\cT_{ (4,3)}$} 
We can use the same numerical algorithm as before to find the MLDE in $\cT_{ (4,3)}$ AD theory. It turns out that  up to the order of $q^{100}$,  the Schur partition function \eqref{T43SchuPT}  satisfies the following  MLDE of   weight 34
  \be
  \cD_q^{(17)}\cZ_{\cT_{ (4,3)}}=0~,
  \ee
  where the explicit form of  MLDO  $  \cD_q^{(17)} $ is given in \eqref{MLDOT43}.
  
  Similarly, we find the indicial equation  
    \be\label{indT43}
  (b-3) (b-1)^2 b^3 (2 b-3)^2 (2 b-1)^3 (4 b-1)^2 g(b)=0~,
  \ee
 where
  \be
  g(b)=b^4-\frac{38 b^3}{3}+\frac{21578712128131 b^2}{344459812152}-\frac{24501532930247 b}{172229906076}+\frac{167773381022507}{1377839248608}~.
  \ee
 The set of solutions is given by   (ignoring multiplicity)
  
    \be
  b=0,\;\frac{1}{4},\;\frac{1}{2},\; 1,\;\frac{3}{2},\;3, \;\beta_1,\;\cdots,  \;\beta_4~,
  \ee
  where $\beta_s$ are the four roots of $g(b)=0$. In particular, note that   $b=0$ is the minimal root and has degeneracy 3.

  \section{High temperature limit  }\label{TpNHighT}
  
  One virtue of modularity is that it relates states in the UV to that in the IR, which allows one to infer the high energy or high temperature behavior. For example, modular invariance of 2D CFT gives rise to the Cardy formula which   characterises  the high energy density of states  universally in terms of central charge \cite{Cardy:1986ie}.
    The same philosophy applies here  for the Schur partition function of 4D $\cN=2$ SCFTs, as we will discuss below.
    
    We would like to understand the behavior of the Schur index $\cI(q) $, or equivalently the Schur   partition function $  \cZ(q)$, in the limit $\tau \to 0$, which will be referred to as the high temperature limit or Cardy limit. 
    
  In many cases, the leading asymptotic of the Schur index / partition function is  \cite{DiPietro:2014bca,Buican:2015ina}
  \be\label{HighT}
  \cI(q)\simeq  \cZ(q)\simeq e^{-\frac{2\pi i }{\tau} 2(a-c)}~, \qquad \tau\to 0~.
  \ee
  It has been observed that this is valid in many example where $a-c<0$, but  violated in a few examples when $a-c>0$. See \cite{ArabiArdehali:2023bpq} for more discussion on this point. The theories studied here have  exactly $a=c$, so the formula \eqref{HighT}, if correct, would predict a finite constant leading term in the limit $\tau\to 0$ or equivalently $q\to 1$.
  
  For $\cN=4$  SYM  theories which   have $a=c$, the Cardy limit of the index has been studied a lot. In particular, 
   the Schur  partition function  of $\cN=4$ $SU(2)$ SYM with central charge $a=c=3/4$ has the leading  asymptotic \cite{ArabiArdehali:2015ybk}
  \be
  \cZ^{\cN=4\; SU(2)\; \text{SYM}}(q) \simeq \frac{1}{-4i\tau}-\frac{1}{2\pi}+\cdots ~,\qquad q\equiv e^{2\pi i \tau}~,
  \ee
  which is obviously different from \eqref{HighT}. This indicates that \eqref{HighT} may be not valid in theories with equal central charges $a=c$. However, the  $\cN=4$  SYM theories  are special as they have enhanced SUSY. 
  
Instead the $\cT_{(p,N)}$ AD theories studied here are  genuine 4D $\cN=2$ SCFTs  with $a=c$. Of course, one can   try to infer the high temperature behavior of Schur index of   $\cT_{(p,N)}$ theories based on  their index relation with $\cN=4$  SYM. However, except for the special case of $p=2$, this requires the knowledge of high temperature limit of flavored Schur index of $\cN=4$  SYM, which is generally not known. \footnote{It would be interesting to use flavored MLDE to understand the high temperature limit of flavored Schur index of $\cN=4$  SYM. The flavored MLDE has been studied e.g. in \cite{Zheng:2022zkm}.}  Given this fact, we will study the high temperature asymptotic behavior    of Schur index of   $\cT_{(p,N)}$ theories directly using MLDE and modular property. In the case of $p=2$, it turns out that we can derive the high temperature  limit for all $\cT_{(2,2k+1)}$ using recursion relation and the defining generating function.  These results motivate us to make some conjectures about MLDE and high temperature  limit of Schur partition function.

 \subsection{$\cT_{(3,2)}$}
Let us first consider the $\cT_{(3,2)}$ AD theory, whose modular properties have been discussed extensively in subsection~\ref{T32MLDE}. We wan to understand the high temperature limit $\tau\to 0$ of Schur partition function $\cZ_{\cT_{(3,2)}}(\tau)$, which is identified with the vacuum character $\chi_1$. 
For our purpose of application, we rewrite \eqref{chi1tau} as
\be
\chi_1(\tau)=\frac{i  }{\sqrt{3} \tau } \Big( \chi_{1}(-\frac1\tau)+\chi_{\frac13}(-\frac1\tau)\Big)+\frac{i}{6 \sqrt{3} \tau }-\frac{1}{6}~.
\ee
This can be used to study the  behavior of the Schur partition function in the high temperature limit, namely
$\tau\to 0, \tilde \tau=-1/\tau\to \infty$, $q\to 1$, $\tilde q=e^{2\pi i \tilde \tau}=e^{ -2\pi i/\tau}\to 0$. In this limit, it is easy to see that $\chi_1(\tilde \tau), \chi_{\frac13}(\tilde \tau) \to 0$ up to exponentially small corrections. Therefore in the high temperature limit $\tau\to0$, we have
\be\label{T32asym}
\cZ_{\cT_{(3,2)}}(\tau)=\chi_1(\tau)=\frac{i}{6 \sqrt{3} \tau }-\frac{1}{6}+O(e^{ -2\pi i/3/\tau})~, \qquad \tau \to 0~,
\ee
up to   exponentially suppressed corrections. 

 \subsection{$\cT_{(2,2k+1)}$}
 
 We now derive the high temperature asymptotic behavior of the Schur partition function  of $\cT_{(2,2k+1)}$ AD theories. We will provide two  ways to derive it. 
 
 The first way to derive is to use  $ \cZ_{\cT_{(2,2k+1)}}(q)= A_k(q^2)$, where  $A_k(q)$ is MacMahon’s generalized ‘sum-of-divisor’ function defined in \eqref{Akq} and satisfies the recursion relation \eqref{Akrecursion} and \eqref{Akq1}. It turns out these formulae are useful enough to derive the asymptotic behavior of $A_k(q)$. 
 
 Let us first derive the asymptotic behavior of $A_1(q)$.  Using the modular behavior of $\bE_2$ in \eqref{E2Es}, we find 
 \be
 \bE_2(\tau) =\frac{1}{\tau^2}\bE_2 (-\frac1\tau) +\frac{1}{2\pi i\tau}~.
 \ee
Meanwhile, we have 
 \be
 \bE_2(\tilde \tau)=-\frac{1}{12} + 2 \tilde q + 6\tilde q^2 + 8\tilde q^3 +\cdots, \qquad \tilde q=e^{2\pi i \tilde \tau} \to 0~,
 \ee
 where $\tilde \tau=-\frac{1}{\tau}\to \infty$. Combining them together, we get
  \be
 \bE_2(\tau) =\frac{1}{\tau^2}\bE_2 (\tilde\tau) +\frac{1}{2\pi i\tau}
 =-\frac{1}{12\tau^2}+\frac{1}{2\pi i\tau} +O(e^{\#/\tau})~, \qquad \tau\to 0~,
 \ee
 up to exponential corrections.  Further using \eqref{Akq1}, we get
  \be\label{A1qasym}
 A_1(q)= \frac12 \mathbb E_{2}(q)+\frac{1}{24 }
  =-\frac{1}{24\tau^2}+\frac{1}{4\pi i\tau}+\frac{1}{24 } +O(e^{\#/\tau})~.
 \ee
 which gives the asymptotic behavior of $A_1(q)$. We can find similar formula for other $A_k(q)$ by using the recursion relation \eqref{Akrecursion}  
  \be\label{Akrecursion2}
 A_k(q) =\frac{1}{2k(2k+1)}\Big[\big(6 A_1(q) +k(k-1) \big) A_{k-1}(q) -2q \frac{d}{dq} A_{k-1}(q) \Big]~.
 \ee
Obviously, we can insert \eqref{A1qasym} into this recursion relation and get the asymptotic behavior of $A_2(q)$. Repeating the procedure in a recursive way, we can find the asymptotic behavior of all $A_k(q)$.
 For simplicity, let us focus on the most singular terms  of $A_k(q)$ in the limit $\tau \to 0$.  By noticing that $A_1 \sim -\frac{1}{24\tau^2}$ and $q\frac{d}{dq}=\tau \frac{d}{d\tau}$, we easily see that the most singular term in \eqref{Akrecursion2} is
   \be\label{Akrecursion3}
 A_k(q) \sim \frac{3}{ k(2k+1)}   A_1(q)  A_{k-1}(q) ~ \sim -\frac{1}{24\tau^2}  \frac{3}{ k(2k+1)} A_{k-1}(q) ~.
 \ee
 With this relation, 
  we can easily  derive the following asymptotic behavior 
 \be
 A_k(q)\sim
 \frac{(-1)^k}{ 8^k k! (2 k+1) !!  }\frac{1}{\tau^{2k}} 
 =
 \frac{(-1)^k}{ 4^k   (2 k+1) !}   \frac{1}{\tau^{2k}}
 \xrightarrow{k\to\infty}
\frac{(-1)^k}{    \sqrt{\pi } 4^{2 k+1} e^{-2 k} k^{2 k+\frac{3}{2}} }\frac{1}{\tau^{2k}}~,\qquad \tau \to 0~,
 \ee
 where we also show the large $k$ limit of the coefficients using Stirling's formula. 
 
 This also gives the high temperature asymptotic behavior of the Schur partition function   
  \be\label{cT2nasy}
 \cZ_{\cT_{(2,2k+1)}}(q=e^{2\pi i \tau})= A_k(q^2)
  =
 \frac{(-1)^k}{ 4^{2 k}   (2 k+1) !}   \frac{1}{\tau^{2k}}
 \xrightarrow{k\to\infty}
\frac{(-1)^k}{    \sqrt{\pi } 4^{3 k+1} e^{-2 k} k^{2 k+\frac{3}{2}} }\frac{1}{\tau^{2k}}~, \qquad \tau\to 0~.
 \ee
 It is very straightforward to  generalize the above derivation and computing all    the subleading corrections. For example, at sub-leading order, we have
 \be
  \cZ_{\cT_{(2,2k+1)}}\sim  \frac{(-1)^k}{ 4^{2 k}   (2 k+1) !}   \frac{1}{\tau^{2k}}
  + \frac{12 i (-1)^k k}{\pi  4^{2k} (2 k+1)!}  \frac{1}{\tau^{2k-1}}+\cdots~,\qquad \tau\to 0~.
 \ee
 
 We now give another derivation of the above asymptotic formula based on the defining generating function \eqref{TpNZNN}, which will be denoted as $\Omega_p(\mu,\tau)$:  
 \be \label{TpNZNN222}
 \Omega_p(\mu,\tau)=
 \sum_{N=1}^\infty \mu^{N-1} (-1)^{\floor{\frac{N}{p}}}  \cZ_{\cT_ {(p,N)}}(q)
 =
 \prod_{j=-\infty\atop j\neq 0}^\infty
  \Big( 1+ \frac{\mu  q^j}{1-q^{j p}}  \Big)
  =\prod_{j=1}^\infty \Big( 1+\frac{\mu  q^j \left(1-q^{j (p-2)}\right)}{1-q^{j p}}-\frac{\mu ^2 q^{j p}}{\left(q^{j p}-1\right)^2} \Big) ~.
 \ee 
 
 In the particular case of $p=2$, we have
 \be \label{ome2mutau}
 \Omega_2(\mu,\tau)=
 \sum_{N=1}^\infty \mu^{N-1} (-1)^{\floor{\frac{N}{p}}}  \cZ_{\cT_ {(2,N)}}(q)
 =1+ \sum_{k=1}^\infty \mu^{2k} (-1)^k  \cZ_{\cT_ {(2,2k+1)}}(q)
  =\prod_{j=1}^\infty \Big( 1 -\frac{\mu ^2 q^{2j }}{\left(q^{2j }-1\right)^2} \Big) ~,
 \ee 
 where we used $ \cZ_{\cT_ {(2,2k)}}=0$ and $ \cZ_{\cT_ {(2,1)}}=1$.  We would like to use this formula to derive the high temperature limit $\tau\to 0$ of $\cZ_{\cT_ {(2,2k+1)}}$. It turns out that in the generating function $\Omega_2$, we need to consider the double scaling limit  by taking  $\mu,\tau \to 0$ but keeping the ratio $\mu/\tau$ fixed.  In such a limit,  the multiplicand in the infinite product of   \eqref{ome2mutau}  reduces to
  \be\label{smalltauint}
 1 -\frac{\mu ^2 q^{2j }}{\left(q^{2j }-1\right)^2} =\left(\frac{\mu ^2}{12}+1\right)+\frac{\mu ^2}{16 \pi ^2 j^2 \tau ^2}+O(\tau^2)
 \xrightarrow{\mu,\tau \to 0,\;\mu/\tau \;\text{fixed}}
 1+\frac{\mu ^2}{16 \pi ^2 j^2 \tau ^2} ~.
 \ee
 Substituting it back to the infinite product, we get \footnote{Rigorously speaking, \eqref{smalltauint}  would fail if $j\tau \gtrsim 1$. However, 
 for big enough $j$, both the LHS and RHS of \eqref{smalltauint}  approach  to 1 due to the suppression of both  $q^{2j },1/j^2\to 0$ in the large $j$ limit.
 }
 \be\label{omeag2muta}
   \Omega_2(\mu,\tau)  =\prod_{j=1}^\infty \Big( 1 -\frac{\mu ^2 q^{2j }}{\left(q^{2j }-1\right)^2} \Big)
    \xrightarrow{\mu,\tau \to 0,\;\mu/\tau \;\text{fixed}}
 \prod_{j=1}^\infty \Big(  1+\frac{\mu ^2}{16 \pi ^2 j^2 \tau ^2}  \Big)  =\frac{4 \tau  }{\mu } \sinh \left(\frac{\mu }{4 \tau }\right)~,
 \ee
 where we used the formula 
 \be\label{sinhfcn}
 \sinh z =z \prod_{n=1}^\infty \Big( 1+\frac{z^2}{(\pi n)^2}\Big)~.
 \ee
 
 One can then perform series expansion in $\mu$ in order to get the asymptotic behavior  of  $ \cZ_{\cT_{(2,2k+1)}}$  in \eqref{cT2nasy}. Equivalently, we will show that \eqref{cT2nasy} gives the same generating function \eqref{omeag2muta}.
 
Indeed, from \eqref{cT2nasy} we have
  \be\label{cT2nasy22}
 \cZ_{\cT_{(2,2k+1)}}(q) 
  =
 \frac{(-1)^k}{ 4^{2 k}   (2 k+1) !}   \frac{1}{\tau^{2k}}+ \frac{\#}{\tau^{2k-1}}+\cdots~,
 \ee
where $  \#$ and dots represent less singular terms and exponentially suppressed terms.  Substituting it back to the generating function  \eqref{ome2mutau}, we get
 \be
  \Omega_2(\mu,\tau) =1+ \sum_{k=1}^\infty \mu^{2k} (-1)^k  \cZ_{\cT_ {(2,2k+1)}}(q)
  \simeq 1+ \sum_{k=1}^\infty \mu^{2k} (-1)^k
 \frac{(-1)^k}{ 4^{2 k}   (2 k+1) !}   \frac{1}{\tau^{2k}} 
 =\frac{4 \tau  }{\mu } \sinh \left(\frac{\mu }{4 \tau }\right)~, \quad
 \ee
 which exactly recovers \eqref{smalltauint}.
Note that in the above formula, we also need to take the double scaling limit $\mu,\tau \to 0$ with $\mu/\tau$ fixed, in order to suppress the contribution from the less singular terms in \eqref{cT2nasy22}.
 
  In the case of $k=1$, \eqref{cT2nasy} gives
  \be
    \cZ^{\cT_{(2,3)}}(q)\sim - \frac{1}{96 \tau^2}~,
  \ee
  which has also been verified numerically.

 \subsection{General case}\label{generalHighT}
To study the high temperature limit of Schur partition function, the key point is to use the modular transformation of characters:
  \be\label{cZSq}
  \cZ(q)=\chi_{\bm 1}(\tau)=\sum_{j}S_{\bm 1 j}\chi_j(-\frac 1 \tau)~,
  \ee
where $\bm 1$ corresponds to the vacuum, and $\chi_j$ are the characters of some modules of the corresponding  VOA, which are also the  solutions to MLDE. Once we understand the behavior of $\chi_j(\tilde \tau)$ in the limit of $\tilde \tau \to \infty $ or equivalently $ \tilde q =e^{2\pi i\tilde \tau}\to 0$, and the modular $S$-matrix, we can establish the high temperature limit of Schur partition function via \eqref{cZSq}.

Using the modular transformation properties,  the authors in \cite{Beem:2017ooy} proposed the following high temperature limit
\be\label{HighT2}
\cZ\sim e^{\frac{\pi i c_{ \text{eff}}}{12\tau}}, \qquad
c_{ \text{eff}}=c_{2d}-24 \min_i h_i=-24\min_i b_i
\ee
where $b=-\frac{c_{2d}}{24}+h$. By comparing \eqref{HighT} and \eqref{HighT2},  it was further proposed   in  \cite{Beem:2017ooy} that
\be\label{ceff}
c_{ \text{eff}}= -24\min_i b_i
=48(c_{ }-a_{ })~.
\ee
From our previous examples of $\cT_{(3,2)}$ and $\cT_{(2,2k+1)}$, we know that the  asymptotic behaviors in both \eqref{HighT} and \eqref{HighT2} are not valid.  However, \eqref{ceff} seems to be still valid. 

In  AD $\cT_{(p,N)}$ theories  considered here, the two central charges are the same, so $ c_{ }-a_{ }=0$. On the other hand,  in the case of $\cT_{(3,2)},\cT_{(2,3)},\cT_{(4,3)}$, by analyzing the indicial equation of MLDE, we do find that $\min_i b_i=0$.

In \cite{ArabiArdehali:2023bpq}, a more careful analysis was done and found that the  following high temperature limit for Schur partition function
\be
\cZ(q) \sim \sum_i  \tilde A _i\; \tilde q ^{b_i} (\log\tilde q)^{d_i-1}+\cdots
\sim \sum_i A_i \tau^{1-d_i } e^{-2\pi i b_i/\tau}+\cdots
\ee
where $d_i$ is the degeneracy of the root  $b_i$ (counting also all the roots that are less than $b_i$ by an integer), and $\tilde q=e^{ 2\pi i \tilde \tau}, \tilde \tau=-1/\tau$.  

Focusing on the most singular term which gives the leading contribution, we get 
\be
\cZ(q)\sim A_j \tau^{1-d_j} e^{-2\pi i b_j/\tau}, \qquad b_j=\min_i b_i~,
\ee
 where $b_j$ is the minimal $b_i$. Depending on the sign of $\min_i b_i$, we can have either exponentially enhanced / suppressed leading contribution, or power law leading asymptotic if $\min_i b_i=0$.
 
If $\min_i b_i>0$, we see $\cZ(q) \to 0$ in the limit $\tau \to 0$, which looks very unlikely as it means an almost perfect cancellation between bosonic and fermionic states.
If $\min_i b_i=-\alpha<0,\alpha>0,\tau=i \beta$, then we have exponentially growing contribution of the form $e^{ 2\pi   \alpha/\beta}$, namely
\be
\log \cZ(q) \sim  \frac{ 2\pi   \alpha}{ \beta}\to +\infty~, \qquad \beta\to 0~.
\ee
This is reminiscent of the index for counting  black hole entropy.~\footnote{Indeed,   the $1/16$-BPS black hole entropy was reproduced from the exponentially large term of    $\cN=4$ SYM   index    in the Cardy-like limit \cite{Choi:2018hmj}.}    Since our theories have $a=c$ which means that they are supposed to   have  nice holographic dual description,   such a kind of exponential   contribution to Schur partition function would indicate the presence of 1/4-BPS black holes.  However, in all our explicit examples, we did not see any case with $\min_i b_i<0$.

On the other hand, if $\min_i b_i=0$, we   have
\be\label{Zgrow}
\cZ(q) \sim \frac{ A_0 }{ \tau^{d_0-1}}~, \qquad \tau\to 0
\ee
where $d_0$ is the degeneracy of the root  $b=0$, namely  in the indicial equation we  have
 the factor $b^{d_0}$. Interestingly, we find that all the examples we studied  fall into this class.
 For $\cT_{(3,2)}$, the asymptotic growth is given by \eqref{T32asym}, while for $\cT_{(2,2k+1)}$, 
 the asymptotic growth \eqref{cT2nasy}.  For  $\cT_{(2,3)}$ and  $\cT_{(4,3)}$, their MLDEs are shown explicitly in the previous section, and their indicial relation are given by \eqref{indT23} \eqref{indT43}. As a result, we do see that the $\min_i b_i=0$ and the degeneracy is 3, namely $d_0=3$ in \eqref{Zgrow}.

Based on these discussions and the results   in the literature, we are then motivated to make the following set of conjectures: 

1) There is no constant term in MLDE, so $\chi=1$ is a solution to MLDE and $b=0$ is a root to the indicial equation;

2)  $b=0$  is the minimal root   to indicial equation; 

3) The degeneracy of root $b=0$  is $N $, so the high temperature asymptotic behavior of the Schur partition function of $ {\cT_{(p,N)}} $ with $\gcd(p,N)=1$ is
 \be\label{ZTpNScaling}
    \cZ^{\cT_{(p,N)}}(q)\sim  \frac{\#}{  \tau^ {N-1}}~, \qquad \tau\to 0~.
  \ee
Note that this kind of asymptotic behavior has been proved for $p=2$.    Moreover  conjecture 1) is a consequence of \eqref{ceff}.
  
  
 To further understand the case of other $p$, let us attempt to generalize the previous generating function techniques.
As in \eqref{smalltauint}, we can take the double scaling limit of the multiplicands in \eqref{TpNZNN222}
  \be
   \Big( 1+\frac{\mu  q^j \left(1-q^{j (p-2)}\right)}{1-q^{j p}}-\frac{\mu ^2 q^{j p}}{\left(q^{j p}-1\right)^2} \Big)
      \xrightarrow{\mu,\tau \to 0,\;\mu/\tau \;\text{fixed}}
      1
      +\frac{\mu ^2}{4 \pi ^2 j^2 p^2 \tau ^2}~,
  \ee
  where the linear term in $\mu$ is absent due to the  double scaling limit. Taking infinite product, we get 
   \beqn
  \Omega_p&=& 
 \prod_{j=1}^\infty    \Big(  1 +\frac{\mu ^2}{4 \pi ^2 j^2 p^2 \tau ^2} \Big)
=
  \frac{2 p \tau   }{\mu }\sinh \left(\frac{\mu }{2 p \tau }\right) ~,
    \eeqn
  where we used \eqref{sinhfcn} again.  This is an even function in $\mu$, so it  would give  zero to $\cZ_{\cT_{(p,N)}}$ for even $N$ according to \eqref{TpNZNN222}. But \eqref{T32asym} has shown that $\cZ_{\cT_{(3,2)}}\sim  \frac{i}{6 \sqrt{3} \tau }$.  This discrepancy is supposed to arise from the order of taking the double scaling   limits and infinite product.
  To exemplify this point, let us consider the  $\mu$-linear  term in \eqref{TpNZNN222}:
    \beqn
  \Omega_p&=& 
 1 +\sum_{j=1}^\infty\frac{\mu  q^j \left(1-q^{j (p-2)}\right)}{1-q^{j p}} +O(\mu^2)
     \\&\sim& \label{Lbdapp}
 1+\sum_{j=1}^\infty\frac{p-2}{p}\mu q^j +O(\mu^2)
     \\&=&
 1+\mu\frac{p-2}{p}  \frac{1}{1-q}+O(\mu^2)
       \\&=&
 1+ \frac{2-p}{p}  \frac{\mu}{2\pi i \tau} +O(\mu^2)~.
  \eeqn
  This would give $\cZ_{\cT_{(3,2)}}\sim  \frac{i}{6  \pi \tau }$, which is  slightly different from but close to  \eqref{T32asym}; the difference  in the coefficients is due to our approximation in \eqref{Lbdapp}. Nevertheless, we get the right scaling behavior consistent with conjecture  \eqref{ZTpNScaling} for $N=2$. A very careful and systematic  analysis is needed to study the general case and get the exact coefficient, and we leave this important question to the future.

\section{Conclusion}\label{conclusions}

In this paper, we studied the Schur sector of a family of AD theories denoted as $\cT_{(p,N)}$. The theories we  have studied are interesting as they share many features with $\cN=4$ SYM theory. In particular, the two central charges are the same  $a=c$, indicating that they have  holographic dual descriptions in terms of supergravity in AdS with some special features. 

We   derived an enlightening  formula  \eqref{TpNSchur} for the Schur index of this family of AD theories, which naturally factors out the Casimir term. The remaining    Schur partition function takes a simple form  \eqref{TpNZNN} and is expected to satisfy the modular linear differential equation. We   study the MLDEs   numerically based on  our simple formula. We also     derive the MLDE analytically and investigate the modular properties of its solution for the theory $\cT_{(3,2)}$, whose VOA is known.  Combining the modularity of MLDE with  the explicit simple formula for Schur partition function, we discuss their high temperature limits. All of our explicit results suggest that the high temperature limit of Schur index  / partition function of $\cT_{(p,N)}$ AD theories diverges following a power law, rather than exponentially. 
This motivates us to propose a set of conjectures about MLDEs and the high temperature behavior \eqref{ZTpNScaling} for general $\cT_{(p,N)}$ AD theories. In the case of $p=2$, we prove the conjecture and show the asymptotic growth explicitly \eqref{ZTpNScaling}. In general, the  exponential growth of index in the high temperature limit is closely related to the black hole entropy in the dual AdS quantum gravity.~\footnote{More precisely, one needs to consider   the large $N$ limit which is different from the high  temperature limit or Cardy limit.  The presence of black holes would give rise to an exponential growth of partition function  in   the large $N$ limit.} Our conjecture on power law divergence \eqref{ZTpNScaling}   indicates the absence of 1/4-BPS black hole.

In the appendix, we also review many important concepts and useful techniques which are used  for developing the results in the main body. In addition, we  also  present  some new   results there, including   computing the torus one-point  function to higher weight     \eqref{STtraceP12}\eqref{STtraceP14},
and deriving     the explicit MLDEs for some AD theories in the family of $(A_{k-1},A_{n-1})$ and $D_p(SU(N))$  in appendix~\ref{MLDEADAA} and \ref{MLDEADDpSUN}, whose chiral algebras are W-algebras and Kac-Moody algebras.

There are some open questions. First of all, it would be interesting to study the properties of the functions  \eqref{TpNZNN}    from a mathematical perspective. In particular, these functions are well-defined for general $p$ and $N$ (some may be trivial), rather than just for  coprime integers $p=2,3,4,6$ and $N=2,3,4\cdots$.
We expect they all enjoy some nice modular properties. For example, one can take $p=4$ and $N=2$ which does not correspond to the SCFT studied here, and check that $\cZ_{ \cT_{(4,2)}}$ defined by \eqref{TpNZNN}   satisfies a MLDE of weight 18. It is also important to study the asymptotic  behavior of these functions in the limit $\tau\to 0$. This characterises the high energy/temperature  growth of 1/4-BPS states in SCFTs.  
We have made the conjecture in  subsection~\ref{generalHighT}, and it would be interesting to prove or disprove it.  Another interesting and related limit is $\tau\to \mathbb Q$.

Secondly, it would be   fascinating to find the chiral algebra of $\cT_{(p,N)}$ AD theory, and use it to  derive the MLDE analytically. So far, only  the chiral algebra of $\cT_{(3,2)}$ AD theory is known. 
 The intriguing operator map between $\cT_{(p,N)}$ AD theory and $\cN=4$ SYM theory proposed in \cite{Buican:2020moo} may offer some  insights and help   to find or bootstrap the chiral algebra.

Thirdly, our Schur partition function is derived from the index of $\cN=4 $ SYM theory based on the index relation  \eqref{TpNIndexrelation}. Then a natural question is   whether there is a direct connection at the level of MLDE. 
It would be amazing if one  could establish an explicit map between the (flavored) MLDE of  $\cN=4 $ SYM theory and the MLDE of $\cT_{(p,N)}$ AD theory.   For $\cN = 4$ $SU(N)$ SYM with odd $N\le 7$ , it has been  observed in   \cite{Beem:2017ooy} that  
  Schur partition functions    satisfy monic MLDEs of weight $   (    N+1 )^2/2 $. 
  It is natural  to ask about the dependence of the weights of MLDEs on the parameters $p$ and $N$ for $\cT_{(p,N)}$ AD theory.

Finally, it would be interesting to find the holographic dual of $\cT_{(p,N)}$ AD theories and study various properties using supergravity techniques. The important feature of $a=c$ implies some remarkable cancellations of   higher derivative  corrections in the supergravity Lagrangian.  Once the holographic dual is known, one could then study the ``giant graviton expansion'' of the Schur index of $\cT_{(p,N)}$ AD theories and try to reproduce it from supergravity. \footnote{In particular,    it is easy to show that,  in the supergravity limit $N\to \infty$, the Schur index is given by 
$\cI_{\cT_{(p,\infty)}}=\prod_{k=1}^\infty\frac{1-q^{k p}}{\left(1-q^k\right) \left(1-q^{k (p-1)}\right)}/\PE\Big[\frac{q^{p-1}-2 q^p+q}{1-q^p}\Big]=\frac{\left(q;q^p\right){}_{\infty } \left(q^{p-1};q^p\right){}_{\infty }}{(q;q)_{\infty } \left(q^{p-1};q^{p-1}\right){}_{\infty } \left(q^p;q^p\right){}_{\infty }}$. }

We leave these interesting questions to the future.

\section*{Acknowledgements}
 We  Matthew Buican, Amihay Hanany and Seyed Morteza Hosseini for related discussions at various stages of this project.
This work  was supported in part by the STFC Consolidated Grants ST/T000791/1 and ST/X000575/1.

\appendix
\section{Modular form and modular linear differential equation} \label{modformMLDE}
We review some basics of modular forms and modular linear differential equations. See e.g. 
\cite{mfs}  for more details.

\subsection{Modular form and Eisenstein series} 
Let $\tau\in \mathbb H$ be the modular parameter taking value in the upper half place. The modular group of interest here is $\Gamma=\PSL(2,\bZ)$. It acts on the modular parameter as 
\be
\tau \to \frac{a\tau+b}{c\tau+d}~, \qquad \gamma=\PBK{a& b\\c & d}\in \Gamma~, \quad a,b,c,d\in \bZ~, \quad ad-bc=1~.
\ee
The modular group is generated by two elements $S$ and $T$
\be
S: \tau \to -\frac{1}{\tau}, \qquad 
T: \tau \to \tau+1
\ee
subject to the relations $S^2=(ST)^3=1$.

For convenience, we define $q\equiv e^{2\pi i \tau}$. The  modular group  action on $q$ is naturally given by  
\be
\gamma \circ q =e^{2\pi i \tau  \frac{a\tau+b}{c\tau+d}}~.
\ee

A modular form of weight $k$ is a holomorphic function $f: \mathbb H \to \bC$ that transforms according to 
\be
f\Big( \frac{a\tau+b}{c\tau+d}\Big) =(c\tau +d)^k f(\tau), \qquad \PBK{a& b\\c & d}\in \Gamma
\ee
and remains finite in the limit  $\Imag  \tau \to +\infty$. Due to the invariance under $T$ transformation, any modular form has a convergent Fourier expansion in $q$ and is finite in the limit $q\to 0$: 
\be
f(\tau) =\sum_{n=0}^\infty a_n q^n~, \qquad q=e^{2\pi i \tau}~.
\ee
One can relax the finiteness condition at infinity and  allow a finite  number of terms with  negative $q$ exponents in the above Fourier expansion, which defines a weakly holomorphic modular form.

A particular set of modular forms is given by the  Eisenstein series  which is defined by \footnote{We will be sloppy about the notation, and use both  $ \mathbb E_{2k}(q) $ and $  \mathbb E_{2k}(\tau) $ to denote the same  Eisenstein series. More generally, we will use $f(q)$ and $f(\tau)$ to denote the same function.}
   \be\label{Einseries}
  \mathbb E_{2k}(\tau) 
  =-\frac{B_{2k}}{(2k)!}+\frac{2}{(2k-1)!}\sum_{n=1}^\infty \frac{n^{2k-1}q^n}{1-q^n}~, \qquad 
  q\equiv e^{2\pi i \tau}~.
  \ee
  where   $B_{ 2k}$ is the $2k$-th Bernoulli number.  When the  integer  $k> 1$, $  \mathbb E_{2k}(\tau)$ is a modular form   for $\Gamma$ of weight $2k$.  The case of $\bE_2(\tau )$ is special as  it is not a
modular form but a quasi-modular form since it transforms anomalously under modular
transformations:
\be\label{E2Es}
\bE_2\Big(\frac{a\tau+b}{c\tau+d} \Big)=(c\tau+d)^2 \bE_2(\tau) -\frac{c(c\tau+d)}{2\pi i}~.
\ee

The space of modular forms of weight $k$ is denoted by $\cM_k(\Gamma)$.  
The ring of the modular forms for modular group $\Gamma=\PSL(2,\bZ)$ is freely generated by $\bE_4(\tau)$ and $\bE_6(\tau)$: 
\be
\bigoplus _{k=0}^\infty  \cM_k (\Gamma) =\bC[E_4(\tau), E_6(\tau)]~.
\ee
In particular, this implies that $\bE_{2k}$ with $k\ge 4$ can be written as the polynomial of  $\bE_4$ and $\bE_6$. For example,
  \be
  \bE_8=\frac37 \bE_4^2~, \qquad 
  \bE_{10}=\frac{5}{11}\bE_4\bE_6~,\qquad \cdots~.
  \ee

The space $\cM_k(\Gamma)$ is finite dimensional:
\be\label{dimG}
\dim  \cM_k(\Gamma)=\begin{cases}
0 ~,& k<0 \text{ or } k \text{ is odd} \\ \vspace{.081cm}
\floor{\frac{k}{12}} ~,& k\equiv 2 \;(\text{mod }12)\\  
\floor{\frac{k}{12}} +1 ~,&  \text{   otherwise} \\
\end{cases}~.
\ee
We can further genearlize and define the vector-valued modular form similarly. Consider a homomorphism 
\be
\rho: \Gamma \to \GL(n, \mathbb C)~,
\ee 
which gives an $n$-dim representation of $\Gamma$. A vector-valued modular form of weight $k \in \bZ$ and multiplier system $\rho$ is a   function   $\bm\chi=\PBK{\chi_0 \\ \vdots \\ \chi_{n-1}}: \mathbb H \to \mathbb C^n$ which transforms as
\be\label{vvmf}
\bm\chi\Big(\frac{a\tau+b}{c\tau+d}\Big) =(c\tau+d)^k \rho(\gamma)\bm \chi(\tau)~, \qquad \gamma \in \Gamma~,
\ee
and all the component functions are finite in the limit $\Imag \tau \to +\infty$.  If  the component functions have exponential growth in this limit, then $\chi(\tau)$ is referred to as a weakly holomorphic vector-valued modular form. 

A weakly holomorphic logarithmic vector-valued modular form of weight $k\in \bZ$ with
multiplier system $\rho$ is    a    function  $\chi: \mathbb H \to \mathbb C^n$ which transforms as \eqref{vvmf} and    such that its $q$ expansion contains logarithms of $q$.

Let us  also introduce the notion of quasi-modular form. The function $f: \mathbb H \to \mathbb C$ is a quasi-modular form of weight $k$ and depth $s$ if there exist holomorphic functions $f_0, \cdots, f_s$ with $f_s\neq 0$, such that
\be
f\Big( \frac{a\tau+b}{c\tau+d}\Big) =(c\tau +d)^k \sum_{j=0}^s f_j(\tau) \Big( \frac{c}{c\tau+d}\Big)^j~, \qquad \PBK{a& b\\c & d}\in \Gamma~, \qquad \tau\in \mathbb H~. 
\ee
We use  $\cM_k^s $ to denote the set of quasi-modular forms of weight $k$ and depth $s$, and  $\cM_k^ {\le s} $ to denote that of weight $k$ and depth non-greater than $s$. Given $f\in \cM_k^s$, we  can always write 
\be
f(\tau)= \sum_{i=0}^s g_i(\tau) \; \bE_2^i(\tau)~, \qquad \exists \; g_i (\tau)\in \cM_{k-2i}~.
\ee
This means 
\be
\cM_k^ {\le s} =\bigoplus_{i=0}^s \cM_{k-2i} \; \bE_2 ^i~.
\ee
So  the ring of  quasi-modular forms is $\mathbb C[\bE_2, \bE_4, \bE_6]$ and the maximal power of $\bE_2$ gives the depth.
 
\subsection{Modular covariant derivative  and MLDE}
The modular forms also transform nicely under the differential operations. For this purpose, we need to introduce the Serre derivative $\p_{(k)}$ which maps modular forms of a fixed weight to   modular forms of higher weight: 
\be\label{serrder}
\p_{(k)}: \cM_k(\Gamma) \to \cM_{k+2}(\Gamma)~.
\ee
More explicitly, Serre derivatives are defined as: 
  \be
  \p_{(k)} f(q)=(q\p_q +k \bE_2 (\tau) ) f(q)~.
  \ee
  When acting on the  Eisenstein series of low-weight, we have 
  \be
    \p_{(4)}\bE_4 =q\p_q \bE_4+4 \bE_2\bE_4=14\bE_6~, \qquad
                     \p_{(6)}\bE_6 =q\p_q \bE_6+6 \bE_2\bE_6=  \frac{60}{7}   \bE_4^2~, \qquad \cdots~,
  \ee
  and
  \be
         \p_{(2)}\bE_2 =q\p_q \bE_2+2\bE_2^2=5\bE_4+\bE_2^2~, \qquad
     \ee
although $\bE_2$ is not a modular form.

Using Serre derivatives we can define $k$-th order modular differential operators  
   \be\label{modcovder}
  D_q^{(k)}=\p_{(2k-2)} \circ\cdots \circ \p_{(2)}\circ \p_{(0)}~,
  \ee
  which  naturally act on objects of modular weight zero, as one can see from \eqref{serrder}. We also define $  D_q^{(0)}f(q):=f(q)$. The modular differential operators   transform with
weight $2k$ under the modular group action, 
  \be
    D_{ \gamma \circ q}^{(k)}=(c\tau+d)^{2k }  D_q^{(k)}~, \qquad \gamma\in \Gamma~.
  \ee
This enables us to construct a large class of modular linear differential operators (MLDO) of a fixed weight $2k$
as sums of modular differential of weight $2k-2r$ multiplied by modular forms of weight $2r$. We are
particularly interested in those that are holomorphic and monic
\be\label{MLDOs}
\cD_q^{(k)}=D_q^{(k)}+\sum_{r=1}^k f_r (\tau) D_q^{(k-r)}~, \qquad f_r (\tau) \in \cM_{2r}(\Gamma)~.
\ee
Here monic means the coefficient of  $D_q^{(k)}$ is one, and by holomorphic we
mean that the modular forms  $f_r (\tau)$ defining the coefficients of the MLDO  are modular forms and hence finite in the limit  $q\to 0$.

The elements of the kernel of a MLDO define a modular linear differential equation (MLDE):
 \be\label{MLDEeq}
\cD_q^{(k)}\chi(\tau)=\Big(D_q^{(k)}+\sum_{r=1}^k f_r (\tau) D_q^{(k-r)} \Big) \chi(\tau)=0~.
\ee
From \eqref{dimG}, one can show  that the total number of undetermined  coefficients in $f_r$  is given by
\be
\dim \sum_{r=1}^k f_r (\tau) D_q^{(k-r)} =\sum_{r=1}^k\dim  \cM_ {2r}(\Gamma)=\left\lfloor \frac{k^2}{12}  +\frac12\right\rfloor +\left\lfloor \frac{k }{2}\right\rfloor~.
\ee
This growth is mild, and allows one to find the MLDE satisfied by $\chi$ numerically with low computational cost. For example, if we want to find a MLDE of weight 60 with $k=30$ above, we need to fix less than 100  coefficients \eqref{MLDEeq}, which means we can determine such a MLDE by just  expanding $\chi$ up to the order around $q^{100+b} $ where $b$ is the smallest power of $q$.

\subsection{Solution  to MLDE}\label{MLDEsol}
Given the MLDE, we would like to find the full set of the solutions, which are supposed to transform as a vector-valued modular form. 

It is easy to see that the MLDO takes the following form 
\be
\cD_q^{(k)}=\sum_{i=0}^k P_{2i}(\tau) q^{k-i}\frac{\p^{k-i}}{\p q^{k-i}}~,
\ee
where $P_{2i}(\tau)$ are polynomials in Einstein series  $\bE_2, \bE_4, \bE_6$.   So it is a Fuchsian ordinary differential equations with only a regular singularity at $q = 0$ inside the unit disk.
 
 To solve MLDE \eqref{MLDEeq}, one can make the following ansatz
 \be
 \chi(\tau) =q^\alpha \sum_{n=0}^{\infty} a_n q^n~,
 \ee
and  substitute it back to \eqref{MLDEeq}, then we get 
 \be\label{mldesol}
 q^{\alpha -k}\Big( \alpha(\alpha-1) \cdots (\alpha -k+1) a_0+\cdots\Big) 
 + q^{\alpha -k+1}\Big( \cdots\Big) 
 +\cdots=0~,
 \ee
 where  the coefficient of  $ q^{\alpha -k}$  is the indicial polynomial with degree $k$. So there are $k$ roots to the indicial polynomial, denoted by $\alpha_1, \cdots, \alpha_k$.

  If all the roots are different and do not differ by an integer, namely $e^{2\pi \alpha_i}$ are all different, then the set of series
 \be
 \chi_i(\tau) =q^{\alpha_i}\phi_i(q)=q^{\alpha_i}\sum_{n=0}^\infty a_{i,n}q^n~, \qquad a_{i,0}\neq 0~,
 \ee 
 is the full set of linearly independent solutions to the MLDE, where the  coefficients $a_{i,n}$ are fixed recursively from   \eqref{mldesol} by  requiring that the coefficients of all $q^\#$  should vanish. They would transform as a vector-valued modular form under the modular group $\Gamma$ \eqref{vvmf}, and   $\rho(T)$ is a diagonal matrix with diagonal elements $e^{2\pi i \alpha_i}$.
 
 On the other hand, if some roots are the same or have an integer difference, then the solution of MLDE may contain logarithmic terms. In this case, the general solution  of a monic holomorphic MLDE of degree $k$ is a $k$-dimensional vector with components  
 \be
\chi_i(\tau) =q^{\alpha_i} \sum_{j=0}^{N_i-1}(\log q)^j\phi_{i,j}(q)~, \qquad \phi_{i,j}(q)=\sum_{l=0}^\infty a_{i,j,l}q^l ~,
 \ee 
 where $N_i$ depends on the structure  of the roots to indicial equation. See  e.g. \cite{ODEsol} for detailed discussions. In particular $N_i\le k$.

The set of linearly independent solutions of MLDE  transform as a weakly holomorphic logarithmic  vector-valued modular form under the modular group $\Gamma$, and   $\rho(T)$ is not a diagonal matrix any more; the off-diagonal entries arise from the logarithmic term due to the fact that $\log q\to \log q+2\pi i $ when we perform $T$-transformation  $\tau \to \tau+1$. 

 \section{Zhu's recursion relation} \label{Zhurecursion}
 We will review   Zhu's recursion relation \cite{zhu} and its application in deriving modular linear differential  equation.   A detailed discussion can be found in the original reference \cite{zhu}  and \cite{Gaberdiel:2008pr,Beem:2017ooy}.
 
 \subsection{Vertex operator algebra}
 We now introduce the vertex operator algebras. We will only introduce the basic concepts and refer the reader to  \cite{Gaberdiel:2008pr,Beem:2017ooy} for details. For simplicity, we will  only consider $\bZ_{\ge 0}$-graded conformal vertex operator algebra, meaning that the vertex operator algebra has a subalgebra which is isomorphic to Virasoro vertex operator algebra, and the conformal dimension of all operators are non-negative integers.  As a result, the vertex operator algebra  $\cV$ can be decomposed as
 \be
 \cV=\bigoplus_{n=0}^\infty \cV_n~, \qquad \dim \cV_n <\infty~,
 \ee
 and $L_0$ operator from Virasoro algebra acts as $L_0a =na$ for $a\in \cV_n$.
 
 The operator-state correspondence dictates that for each state $a$ with integer weight $h_a$, the corresponding mode expansion of the corresponding vertex operator is given by 
 \be\label{amodexp}
 a(z) =Y(a,z) =\sum_{n\in \bZ}a_{-h_a -n }z^n
  =\sum_{n\in \bZ}a_{n}z^{-h_a -n}~,
 \ee
 where  the modes act as endomorphism $a_{n}: \cV_k \to \cV_{k+n}$.  
 
 We   further introduce the following notation for the zero mode of $a$ 
\be
o(a) := a_0~.
\ee
 
For the convenience of    formulating recursion relations on the torus, we consider an alternative expansion of the same vertex operator 
  \be
Y[a,z] =e^{zh_a}Y(a,e^z-1)=\sum_{n\in \bZ}a_{[-h_a -n]}z^n~,
 \ee
 where we have introduced the ``square-bracket'' modes. They are related to the usual modes in  \eqref{amodexp} via
 \be
 a_{[n]}=\sum_{j\ge n} c(j+h_a-1,n+h_a-1;h_a) a_{ j }~,
 \ee
 where the coefficients are defined via
 \be  \label{sqmode}
 (1+z)^{h-1}\Big(\log(1+z)\Big)^n=\sum_{j\ge n}c(j,n;h) z^j~.
 \ee
 For Virasoro primary state $a$, the commutation relations for  $a_{[n]}$ are identical to those of $a_{ n }$. 
 For stress tensor, which is not   Virasoro primary, we instead define
 \be
L_{[n]}=\sum_{j\ge n} c(j+ 1,n+ 1;h_a)  L_{j} -\frac{c}{24}\delta_{n+2,0}  ~.
 \ee
 These square-bracket Virasoro modes then satisfy the same commutation relation as the original Virasoro modes $L_n$.  
 
 We also need the vacuum state which is defined such that \footnote{With this property of vacuum, we can define a state $\ket{a} =\lim_{z\to 0}a(z) \ket{\Omega} =a_{-h_a}\ket{\Omega}$.}
  \be\label{vacuum}
 a_{n}\Omega=a_{[n]}\Omega=0~, \qquad n \ge -h_a~.
 \ee
 The two definitions of vacuum are consistent  as   $a_{[n]}=a_{n}+c(n+1,n;h_a)a_{ n+1 }+\cdots$ which can be shown  from \eqref{sqmode}.
 
Since the vacuum state and the commutation relations defined via the square-bracket modes are the same as those of  usual modes,  a  null vector in the vacuum
Verma module of a VOA formulated in terms of the  usual mode will still be   null after replacing the usual modes with square bracket modes. In another word, once we have a null state acting on vacuum, we can replace   all the $a_n$ in the null state with $a_{[n]}$, and the resulting state is still null.
 
  \subsection{Zhu's recursion relation   for torus one-point function }
  
  The virtue of the square-bracket modes is that one can elegantly formulate the recursion relation  for  torus one-point functions  in the VOA. We will call such   kinds of recursion relation as Zhu's  recursion relation, as they were originally derived by Zhu \cite{zhu}. The  simplest   Zhu's  recursion relation is   \footnote{
  Here we consider the general case of vertex operator superalgebras, which have even and odd parts  $\cV=\cV_{ \text{even}}\oplus \cV_{ \text{odd}}$. The supertrace of an endomorphsim $\cO: \cV\to \cV$ is defined as $ \STr_\cV\equiv \Tr_{\cV_{ \text{even}}}\cO- \Tr_{\cV_{ \text{odd}}}\cO$.
}
 \be\label{Zhu}
 \STr_\cV \Big(o\big(a_{[-h_a]}b\big)q^{L_0-\frac{c}{24}}\Big)
 = \Tr \Big(o(a)o(b)q^{L_0-\frac{c}{24}}\Big)
 + \sum_{k\ge 1}\bE_{2k}(\tau)\Tr \Big(o\big(a_{[-h_a+2k]}b\big)q^{L_0-\frac{c}{24}}\Big)~.
 \ee

 More generally,  for $n\ge 1$ we  have 
 \be\label{Zhu2}
 \STr_\cV \Big(o\big(a_{[-h_a-n]}b\big)q^{L_0-\frac{c}{24}}\Big)
 =  (-1)^n \sum_{2k\ge n+ 1}^\prime \PBK{2k-1\\n}\bE_{2k}(\tau)\Tr \Big(o\big(a_{[-h_a-n+2k]}b\big)q^{L_0-\frac{c}{24}}\Big)~,
 \ee
 where the prime indicates that the term with $ 2k=n+1$ should be removed because $o\big(a_{[-h_a+1]}b\big)$ is a commutator and hence $\STr \Big(o\big(a_{[-h_a+1]}b\big)q^{L_0-\frac{c}{24}}\Big)=0$.  Note that since on the RHS, we have $-h_a-n+2k \ge -h_a+1$, so $a_{[-h_a-n+2k]} $ annihilates the vacuum \eqref{vacuum}. 
After using the commutation relation of the algebra, we can move the left-most oscillator $a_{[-h_a-n+2k]} $  to the right inside $o\big(a_{[-h_a-n+2k]}b\big)$ and obtain the   zero mode of  state with strictly smaller  conformal dimension. This then gives a way to compute the torus one-point function recursively. 

  \subsection{MLDE from null state}
  
  Given a VOA $\cV$, one can define a subspace $C_2(\cV)\subset \cV$ as
  \be\label{C2V}
  C_2(\cV) :=\text{Span} \Big\{a_{-h_a-1}b \Big| a, b\in \cV  \Big\}~.
   \ee
  By operator-state correspondence, the state $ a_{-h_a-1}\Omega$  is associated to the vertex operator $\p^n a$.  The vector space $C_2(\cV)$  can be understood as  the space of those normally-ordered composite operators that contain
at least one derivative, including those operators without explicit derivatives apparently but can be rewritten in terms of  operators  with derivatives after using appropriate null relations in the VOA.  

We can further define the quotient space
\be
\cR_\cV:=\cV/C_2(\cV)~.
\ee
The corresponding algebra is known as the $C_2$-algebra of $\cV$, which is a commutative, associative Poisson algebra.

In many interesting cases, the VOA   contains some null state of the form

\be\label{nullT}
\cN_T=(L_{-2})^k\Omega +\sum_i a_{-h_i-1}^i \varphi_i~.
\ee 
In other words, we have $(L_{-2})^k\Omega  \in C_2(\cV)$ for $k \in \bZ_+$. The existence of such a null state is closely related to the existence of MLDE for the vacuum character    of $\cV$. 

Due to the isomorphism between square modes and usual modes that we discussed before, a null state of the form in \eqref{nullT} indicates another null state in the Verma module of VOA
\be
\cN_{[T]}=(L_{[-2]})^k\Omega+\varphi= (L_{[-2]})^k\Omega+ \sum_i a_{[-h_i-1]}^i b_i~, \qquad
\varphi\in C_{[2]}(\cV)~,
\ee
where $C_{[2]}(\cV)$ is the square mode analogue of \eqref{C2V}.
Because $\cN_{[T]}$ is a null state, correlation functions with  insertions of  $\cN_{[T]}$ must vanish. In particular, the torus one-point function of $\cN_{[T]}$ must be zero
\be
\STr_\cV \Big(o(\cN_{[T]}) q^{L_0-\frac{c}{24}}\Big)
=\STr_\cV \Big(o((L_{[-2]})^k\Omega) q^{L_0-\frac{c}{24}}\Big)+\sum_i \STr_\cV \Big(o( a_{[-h_i-1]}^i b_i )q^{L_0-\frac{c}{24}}\Big)  =0~.
\ee
The torus one-point function with insertions above can be evaluated using Zhu's recursion relation in  \eqref{Zhu} and \eqref{Zhu2}. Consequently and in ideal cases, \footnote{After using the recursion relation, one may encounter the zero mode of the form $o(a^1_{-h_1}\cdots a^j_{-h_j})$, which generally gives an obstruction for further evaluation. See \cite{Beem:2017ooy} for discussions.  We don't encounter such kinds of obstructions in this paper. } we get a modular  covariant differential operator acting on  the vacuum  character, namely we obtain the MLDE for the vacuum  character of VOA \eqref{MLDEeq}
\be
\cD_q^{(k)}\chi(\tau) =\Big(D_q^{(k)}+\sum_{r=1}^k f_r (\tau) D_q^{(k-r)} \Big) \chi(\tau) =0~,\qquad \chi(\tau) =\STr_\cV \Big( q^{L_0-\frac{c}{24}}\Big)~.
\ee
From these  discussions, we see that the  MLDE for the vacuum  character arises as the consequence of  a null state of the form \eqref{nullT}.

\subsection{Stress tensor one-point function }
To derive the  MLDE, one important ingredient is to derive the torus one-point function with insertion $o((L_{[-2]})^k\Omega)$. Let us start with the simplest case of $k=1$.  
Using Zhu's recursion relation \eqref{Zhu}, we get
 \be\label{}
 \STr_\cV \Big(o\big(L_{[-2]}\Omega\big)q^{L_0-\frac{c}{24}}\Big)
 = \STr_\cV \Big(o(L)o(\Omega)q^{L_0-\frac{c}{24}}\Big)
 = \STr_\cV \Big( (  L_0-c/24)q^{L_0-\frac{c}{24}}\Big)
  =q\frac{d}{dq} \STr_\cV \Big( q^{L_0-\frac{c}{24}}\Big)~,
  \ee
  where we used that $L_{[i]}\Omega=0$ for $i>-2$ and $o\big(L_{[-2]}\Omega\big) =o(L)=L_{0}-\frac{c}{24} $.
 We also note that  $
 o(L_{[-1]}a)= o(L_{ -1 }a+L_{0}a)=(L_{ -1 }a+L_{ 0 }a)_{ 0 }
 =-(h_a+0) a_{ 0 }+h_a a_{ 0 }=0
$
 where $L_{[-1]}=L_{ -1 }+L_{ 0 }$.
 
 For higher $k$, we can similarly derive the torus one-point function using   \eqref{Zhu} and \eqref{Zhu2}.
  \beqn
&&
 \STr_\cV \Big(o\big(L_{[-2]}(L_{[-2]})^r\Omega\big)q^{L_0-\frac{c}{24}}\Big)
  \\ &=&
   \STr_\cV \Big(o(L)o((L_{[-2]})^r\Omega)q^{L_0-\frac{c}{24}}\Big)
    + \bE_{2}(q) \STr_\cV \Big(o\big(L_{[0]} (L_{[-2]})^r\Omega\big)q^{L_0-\frac{c}{24}}\Big)
        \nonumber     \\ & & + \sum_{k\ge 2}\bE_{2k}(q)\;\STr_\cV \Big(o\big(L_{[-2+2k]} (L_{[-2]})^r\Omega\big)q^{L_0-\frac{c}{24}}\Big)
          \nonumber     \\ &=&
   \STr_\cV \Big( (L_0-\frac{c}{24})o((L_{[-2]})^r\Omega)q^{L_0-\frac{c}{24}}\Big)
    +  2r\bE_{2}\; \STr_\cV \Big(o\big( (L_{[-2]})^r\Omega\big)q^{L_0-\frac{c}{24}}\Big)
        \nonumber     \\ & & + \sum_{k\ge 2}\;\bE_{2k} \;\STr_\cV \Big(o\big(L_{[-2+2k]} (L_{[-2]})^r\Omega\big)q^{L_0-\frac{c}{24}}\Big)
       \nonumber     \\ &=&
 \Big( q\frac{\p}{\p q} +  2r  \bE_{2} \Big)\STr_\cV \Big(o\big( (L_{[-2]})^r\Omega\big)q^{L_0-\frac{c}{24}}\Big)
     + \sum_{k\ge 2}\bE_{2k}\; \STr_\cV \Big(o\big(L_{[-2+2k]} (L_{[-2]})^r\Omega\big)q^{L_0-\frac{c}{24}}\Big) 
         \nonumber             \\ &=&
\p^{(2r)}\STr_\cV \Big(o\big( (L_{[-2]})^r\Omega\big)q^{L_0-\frac{c}{24}}\Big)
      + \sum_{k\ge 2}A_{k,r}\bE_{2k} \STr_\cV \Big(o \big(  (L_{[-2]})^{r-k+1}\Omega  \big)q^{L_0-\frac{c}{24}}\Big)~,
  \eeqn
 where  $A_{k,r}$ is defined by
 \be
 L_{[-2+2k]} (L_{[-2]})^r\Omega=A_{k,r}(L_{[-2]})^{r-k+1}\Omega ~.
 \ee
 Explicitly, one can compute $A_{k,r}$ using Virasoro   commutation relations repeatedly and the property of vacuum $L_{[m]}\Omega=0$ for $m>-2$. 
 In particular, one can easily show that $A_{1,r}=2r$, namely $ L_{[0]} (L_{[-2]})^r\Omega= 2r (L_{[-2]})^r\Omega$.  
  
 As a result, we find the following trace formula 
 \be\label{STtraceformula} 
  \STr_\cV \Big(o\big( (L_{[-2]})^r\Omega\big)q^{L_0-\frac{c}{24}}\Big)
=\cP_{2r} \circ \STr_\cV \Big(o\big(  q^{L_0-\frac{c}{24}}\Big)~,
 \ee
 where
 \beqn  
  \cP_2 &=&D_q^{(1)}~,  \label{STtraceP2}
 \\
 \cP_4 &=&D_q^{(2)}+\frac{c  }{2}\mathbb{E}_4~, \label{STtraceP4}
 \\
  \cP_6 &=&D_q^{(3)}+2  \left(4+\frac{3 c}{4}\right) \mathbb{E}_4 D_q^{(1)}+10 c \mathbb{E}_6~,  \label{STtraceP6}
 \\
  \cP_8 &=&D_q^{(4)}+(32+3 c) \mathbb{E}_4D_q^{(2)}+40  (4+c) \mathbb{E}_6 \;D_q^{(1)}+\frac{3}{4} c (144+c) \mathbb{E}_4^2 ~, \label{STtraceP8}
 \\
\cP_{10}&=&D_q^{(5)}+5 \left((16+c) \mathbb{E}_4\right)D_q^{(3)}
+100  (8+c) \mathbb{E}_6\;D_q^{(2)}
+\frac{5}{4}  (1536+c (464+3 c)) \mathbb{E}_4^2 \; D_q^{(1)} \nonumber\\&&
+\frac{50}{11} c (816+11 c) \mathbb{E}_4 \mathbb{E}_6~,   \label{STtraceP10}
\\
\cP_{12}&=&
D_q^{(6)}+\frac{5}{2}  (64+3 c) \mathbb{E}_4\;D_q^{(4)}
+200 \left((12+c) \mathbb{E}_6\right)D_q^{(3)}
+20  \left(608+93 c+\frac{9 c^2}{16}\right) \mathbb{E}_4^2 D_q^{(2)}
\nonumber\\&&
+\frac{100}{11}  (7936+c (2756+33 c)) \mathbb{E}_4 \mathbb{E}_6\;D_q^{(1)}+\frac{15}{104} c (285696+13 c (432+c)) \mathbb{E}_4^3
\nonumber\\&&
+\frac{1000}{13} c (744+13 c) \mathbb{E}_6^2~, \quad \label{STtraceP12}
\\ 
\cP_{14}&=&
D_q^{(7)}+\frac{7}{2}   (80+3 c) \mathbb{E}_4\; D_q^{(5)}
+350  (16+c) \mathbb{E}_6\;D_q^{(4)}
+\frac{35}{4}  (5120+3 c (176+c)) \mathbb{E}_4^2\;D_q^{(3)}
\nonumber\\&&
+\frac{350}{11}  (17280+c (3064+33 c)) \mathbb{E}_4 \mathbb{E}_6\;D_q^{(2)}
+\Big(\frac{15}{104} (6062080+7 c (335616+13 c (448+c))) \mathbb{E}_4^3
\nonumber\\&&
+\frac{7000}{13} (2176+c (848+13 c)) \mathbb{E}_6^2\Big)D_q^{(1)}
+\frac{525}{22} c (240+c) (576+11 c) \mathbb{E}_4^2 \mathbb{E}_6~,  \label{STtraceP14}
\\ 
\cP_{16}&=&
D_q^{(8)}+14 \left((32+c) \mathbb{E}_4\right)D_q^{(6)}
+560 \left((20+c) \mathbb{E}_6\right)D_q^{(5)}
+\frac{35}{2}  (7168+c (560+3 c)) \mathbb{E}_4^2\;D_q^{(4)}
\nonumber\\&&
+\frac{2800}{11}  (9344+c (1124+11 c)) \mathbb{E}_4 \mathbb{E}_6\;D_q^{(3)}
+\Big(\frac{5}{26} (40845312+7 c (1163264+39 c (464+c))) \mathbb{E}_4^3
\nonumber\\&&
+\frac{28000}{13} (4768+c (952+13 c)) \mathbb{E}_6^2\Big)D_q^{(2)}
+\frac{700}{11}  (1155072+c (492608+c (10132+33 c))) \mathbb{E}_4^2 \mathbb{E}_6\;D_q^{(1)}
\nonumber\\&&
+\frac{105 c (1207885824+17 c (1951488+13 c (864+c)))  }{3536}\mathbb{E}_4^4
\nonumber\\&&
+\frac{14000 c (18873216+17 c (29400+143 c))  }{2431}\mathbb{E}_4 \mathbb{E}_6^2
~,  \label{STtraceP16}
 \eeqn
 where $c$ is the central charge of Virasoro algebra (not the 4d central charge). The expressions of $\cP_{2k}$ for larger $k$ can also be obtained easily, but we don't write them down explicitly here as they are becoming more and more complicated. Note that $\cP_{2k}$'s  up to $k=5$ were  presented in \cite{Gaberdiel:2008pr}.
 
 The same method can be used to derive the general torus one-point function with any Virasoro insertion $o(L_{-n_1}\cdots L_{-n_s}\Omega)$, based on Zhu's recursion relation  \eqref{Zhu}   \eqref{Zhu2} and the Virasoro commutation relation. A particularly simple case is   
 \be\label{oneptfcn1T}
  \STr_\cV \Big(o\big(L_{[ -n]}\Omega\big)q^{L_0-\frac{c}{24}}\Big)=0~, \qquad n>2~,
 \ee
 which can be understood from the fact that   
  the resulting modes $ L_{[-h_a-n+2k]}\Omega$,  
arising from applying the recursion relation  \eqref{Zhu2},  annihilate  the vacuum.
 
 Before closing this subsection, we would like to recall some useful formulae for operator modes and OPEs in order to facilitate the actual derivation  of MLDE in    general VOAs with  generators besides the Virasoro ones.

 Given the mode expansion in \eqref{amodexp}, one can  take derivatives  and  find that modes for the derivative of vertex operator is related to the modes of   original vertex operator  via
   \be\label{dermode}
 (\p a)_m =-(h_a+m) a_m~, \qquad
 (\p^ka)_m=(-1)^k (h_a+m+k-1)\cdots(h_a+m) a_m~.
 \ee

 In physics, one can  concretely  specify  the VOA via the OPEs of   the generators
 \be\label{OPEAB}
 A(z) B(0)\sim \sum_n   \frac{ [AB]_n }{ z^n}~. 
 \ee
In particular, the $n=0$ term defines   the normal order product $:AB: \equiv [AB]_0$ which will also be abbreviated as $AB$ for simplicity. The  modes of the normal order product are given by
 \be
 (AB)_n =\sum_{k\le -h_A}A_k B_{n-k}+(-1)^{|A| |B|}\sum_{k> -h_A} B_{n-k}A_k ~,
 \ee
 where $(-1)^{|A| |B|}$ takes into account the statistics of two operators.  We then have $\ket{(AB)}=(AB)_{-h_A -h_B}\ket{0}=A_{-h_A}B_{-h_B}\ket{0}$. Furthermore, we have $
 A_n B = [AB]_{n+h_A}$.

From OPE in \eqref{OPEAB}, one can then obtain the commutation relation of their modes  \cite{Thielemans:1994er}
 \be
 [A_m, B_n]=\sum_{l>0}\PBK{m+h_A-1\\ l-1} ([AB]_{l})_{m+n}~.
 \ee
 
 Finally, the commutation relation between Virasoro modes and modes of primary operator is given by
 \be\label{LOcom}
 [L_m, O_n]=((h-1) m -n) O_{m+n}~.
 \ee
\section{MLDE for families of AD theories}\label{MLDEAD}
\subsection{$(A_{k-1},A_{n-1})$   }\label{MLDEADAA}

Let us consider the    AD theory   $(A_{k-1},A_{n-1})$ subject to the  coprime condition that $\gcd(k,n)=1$. This family of theory is simple as the chiral algebra is the   W-algebra minimal model
$\cW(k, k + n)$ \cite{Cordova:2015nma}.

The central charges of the corresponding 2d VOAs are
\be
c_{2d} ^{(A_{k-1},A_{n-1})}=-12c_{4d} ^{(A_{k-1},A_{n-1})}=-\frac{(k-1) (n-1) (k+n+nk)}{n+k}~,
\ee
The Schur index is given by \cite{Song:2015wta,Song:2017oew}
\be
\cI_{(A_{k-1},A_{n-1})}=\PE\Big[\frac{ q^2(1-q^{k-1})(1-q^{n-1})}{(1-q)^2 (1-q^{k+n})} \Big]~.
\ee

In the special case of $k=2, n=2r+1$, we get $(A_1, A_{2r})$ AD theory, whose chiral algebra is simply given by $(2, 2r+3)$ minimal model VOA, whose only strong generator is the stress tensor.  The SCFT/VOA correspondence then implies the vacuum character of $(2, 2r+3)$  VOA coincides with the Schur index of  $(A_1, A_{2r})$ up to the Casimir factor, namely
\beqn\label{A1A2rcharacterr}
&&\chi_{(2,2r+3)} \equiv  \STr_{ \cV_{(2,2r+3)}}     q^{L_0-\frac{c}{24}} 
\\&=&\cZ_{(A_1,A_{2r})}=
q^{-\frac{c_{2d}^{(A_{ 1},A_{2r})}}{24}}\cI_{(A_1,A_{2r})}
=q^{\frac{r (6 r+5)}{12(2 r+3 )}} \PE \Big[  \frac{q^2 \left(1-q^{2 r} \right)}{( 1-q) \left(1-q^{2 r+3} \right)}\Big]~.
\eeqn

 {\bf  $(A_1, A_{2 })$ AD theory. }
The VOA of $(A_1, A_{2 })$ AD theory is given by the (2,5) Lee-Yang VOA. It has a null operator of the form
\be
 T^2-\frac{3}{10}\p^2 T=0~.
 \ee
 Equivalently, there is a null state of the form
 \be
 \cN_4\equiv(L_{-4}-\frac53 L_{-2}^2)\Omega=0~.
 \ee

Following the previous general discussions, the torus one-point function with the insertion of the zero mode this  null state is
\beqn
\STr_\cV \Big(o  (\cN_4)  q^{L_0-\frac{c_{2d}}{24}}\Big)
&=&\STr_\cV \Big(o(  L_{[-4]} \Omega  ) q^{L_0-\frac{c_{2d}}{24}}\Big)
 -\frac53\STr_\cV \Big(o( (  L_{[-2]})^2 \Omega) q^{L_0-\frac{c_{2d}}{24}}\Big)
\\&=&
 -\frac53 \cP_{4}\circ
 \STr_\cV \Big(   q^{L_0-\frac{c}{24}}\Big)
 =0~,
\eeqn
 where we have used \eqref{oneptfcn1T} and  \eqref{STtraceformula}. Using \eqref{STtraceP12}, we then find the  following MLDE  
 \be
 \Big(D_q^{(2)}-\frac{11}{5}\bE_4 \Big) \chi_{(2,5)}=0~.
 \ee

 {\bf  $(A_1, A_{4 })$ AD theory. }
 The VOA of $(A_1, A_{4 })$ AD theory is given by the (2,7) minimal model VOA. It has a null operator of the form
 \be
  T^3-\frac{11}{14}  T'' T -\frac{1}{7}  T' T'  -\frac{19  }{588}T^{(4)}=0~.
 \ee
  Equivalently, there is a null state of the form
 \be\label{cN6A1A4}
\cN_6=\Big(L_{-6} +\frac{77}{38}L_{-4}L_{-2} +\frac{7}{38}L_{-3}L_{-3} -\frac{49}{38}L_{-2}L_{-2}L_{-2}\Big) \Omega=0 ~.
 \ee
 Following the general discussions in the previous subsections, the insertion of zero mode of this null state in the torus  partition function leads to a MLDE. In particular, the insertion of the first and last mode in the bracket of \eqref{cN6A1A4} haven been computed in \eqref{oneptfcn1T} and  \eqref{STtraceformula}, and the insertion of the rest of modes can be computed using Zhu's recursion relation  \eqref{Zhu}   \eqref{Zhu2}. Consequently, we find the  following MLDE    
  \be
 \Big(D_q^{(3)}-\frac{100}{7}\bE_4 D_q^{(1)}-\frac{1700}{49}\bE_6 \Big) \chi_{(2,7)}=0~.
 \ee

 {\bf  $(A_1, A_{6 })$ AD theory. }
    The VOA of $(A_1, A_{6 })$ AD theory is given by the (2,9) minimal model VOA, which has a null state of the form 
  \be
\cN_8= \Big(L_{-2}^4-\frac{26}{9} L_{-4} L_{-2}^2-\frac{4}{9} L_{-3}^2 L_{-2}-\frac{88}{27} L_{-6} L_{-2}+\frac{7 L_{-4}^2}{9}+\frac{4}{27} L_{-5} L_{-3}-\frac{278 L_{-8}}{81}\Big)\Omega=0~.
  \ee
 
 As before, this null state gives rise to the following   MLDE  
    \be
 \Big(D_q^{(4)}-\frac{130}{3}\bE_4 D_q^{(2)}-\frac{7420}{27}\bE_6  D_q^{(1)}-\frac{6325}{27}\bE_4^2 \Big)\chi_{(2,9)}=0~.
 \ee

 {\bf  $(A_1, A_{8 })$ AD theory. }    The VOA of $(A_1, A_{8 })$ AD theory is given by the (2,11) minimal model VOA, which has a null state of the form 
 \beqn
 \cN_{10}&=& \Big(L_{-2}^5-\frac{50}{11} L_{-4} L_{-2}^3-\frac{10}{11} L_{-3}^2 L_{-2}^2-\frac{1004}{121} L_{-6} L_{-2}^2+\frac{411}{121} L_{-4}^2 L_{-2}
 \\&&
 +\frac{52}{121} L_{-5} L_{-3} L_{-2}-\frac{22914 L_{-8} L_{-2}}{1331}+\frac{6}{11} L_{-4} L_{-3}^2+\frac{5052 L_{-6} L_{-4}}{1331}
 \\&&
 +\frac{984 L_{-7} L_{-3}}{1331}-\frac{164 L_{-5}^2}{1331}-\frac{205200 L_{-10}}{14641} \Big) \Omega
 =0~.
 \eeqn
 
  As before, this null state gives rise to the following   MLDE  
   \be
 \Big(
 D_q^{(5)} -\frac{1060}{11} \bE_4 D_q^{(3)}-\frac{123900}{121} \bE_6 D_q^{(2)}-\frac{2460400  }{1331} \bE_4^2 D_q^{(1)}-\frac{706764800  }{161051}\bE_4 \bE_6
  \Big)\chi_{(2,11)}=0~.
 \ee
 One can also verify this equation numerically as the character is known \eqref{A1A2rcharacterr}.
 \bigskip

 The case of $k=3$ is also simple, and gives $(A_2, A_{n-1 })$ AD theories.  If $\gcd(n,3)=1$, we have $(A_2, A_{3r }),(A_2, A_{3r+1 })$ AD theories, whose   VOAs are $W_3$-algebra, with strong generators $T$ and $W$.
  
  The OPEs are given by
  \beqn
  T(z)T (0)  &\sim& \frac{c/2}{z^4}+\frac{2T}{z^2}+\frac{T'}{z} ~,\\
T(z)W (0)  &\sim&  \frac{3W}{z^2}+\frac{W'}{z} ~, \\
W(z) W (0)  &\sim&  \frac{c/3}{z^6}+  \frac{2T}{z^4}+\frac{T'}{z^3}+\frac{\frac{3}{10}T''+\frac{32}{22+5c}\Lambda}{z^2}
+\frac{\frac{1}{15}T'''+ \frac{16}{22+5c}\Lambda'}{z}~,
  \eeqn
 where $\Lambda$ is given by \eqref{lmda2}.
 
 In terms of modes, the commutation relations are
  \beqn
[L_m,L_n]&=& (m-n)L_{m+n}+\frac{c}{12}(m^3-m)\delta_{m+n,0}~,
\\ 
{} [L_m,W_n]&=& (2m-n)W_{m+n} ~,
\\ \label{WWcomm}
{} [W_m,W_n]&=& ( m-n)\Big[ \frac{1}{15}(m+n+3)(m+n+2)-\frac16 (m+2)(n+2)\Big] L_{m+n}  
\\&&
+\frac{16}{22+5c}(m-n)\Lambda_{m+n}+\frac{c}{360}m(m^2-1)(m^2-4)\delta_{m+n,0}~.
  \eeqn

    {\bf    $(A_2, A_{3  })$ AD theory. }  The VOA is $W_3$ algebra with $c=-114/7 $.
  At dimension 5, we find a  null operator 
  \be
     N_5=  W''-\frac{14}{3}TW=0~, \qquad
  \ee
    At dimension 6, we find a  null operator 
    \be
   N_6=   168 \left(T'\right)^2+336   T'' T-196 T^3-17 T^{(4)}+1092 W^2=0~.
    \ee
   Then we can consider the normal order product $TN_6$, where $TW^2$  becomes $W''W$ using the null operator $N_5$. Consequently, we find a null operator at dimension 8: 
  \be
  T^4-\frac{6}{7}   \left(T'\right)^2 T-\frac{12  }{7}  T'' T^2+\frac{9  }{28}\left(T''\right)^2+\frac{78  }{49} W''W+\frac{1}{7} T^{(3)}T'  -\frac{9}{98}   T^{(4)} T-\frac{3  }{490}T^{(6)}~.
  \ee
  The corresponding state is
  \beqn
  &&
\Big(L_{-2}^4-\frac{24}{7} L_{-4} L_{-2}^2-\frac{6}{7} L_{-3}^2 L_{-2}-\frac{108}{49} L_{-6} L_{-2}+\frac{9 L_{-4}^2}{7}
\nonumber\\&&\qquad 
+\frac{6}{7} L_{-5} L_{-3}-\frac{216 L_{-8}}{49}+\frac{156}{49} W_{ -5} W_{ -3 }\Big)\Omega=0~.
  \eeqn
  Obviously, we have $L_{-2}^4\in C_2(\cV)$.  We can then use Zhu's recursion relation to derive the MLDE. 
The only non-trivial computation is for $W$ generators, which can be computed using \eqref{WWcomm}:
  \beqn
  W_{m} W_{ -3 } \Omega&=& [W_{m}, W_{ -3 }] \Omega
  \\&=& \nonumber
  \Big[
  \frac{1}{156} (m+3) \Big(\left(23 m^2+3 m+52\right) L_{ m-3}-42 (T^2)_{ m-3}\Big)-\frac{19}{420} m \left(m^4-5 m^2+4\right) \delta _{0,m-3}\Big]\Omega~,
  \eeqn
  which is valid for $m>-3$, as  $W_m\Omega=0$. Here  $(T^2)_{  -4}\Omega=L_{-2}^2\Omega$ and $(T^2)_{  m-3}\Omega=0$ for $m>- 1$.
 
  As a result, we find the following MLDE
  \be
\Big(D_q^{(4)}-\frac{370}{7} \mathbb{E}_4D_q^{(2)}-\frac{10340}{49} \mathbb{E}_6D_q^{(1)}-\frac{115425  }{343} \mathbb{E}_4^2\Big) \cZ_{(A_2,A_3)}=0~.
  \ee
  
 {\bf    $(A_2, A_{4  })$ AD theory. } The VOA is $W_3$ algebra with   $c=-23$. The MLDE can be derived similarly and turns out to be given by
  \be
\Big(D_q^{(5)}-125 \mathbb{E}_4D_q^{(3)}-\frac{24125}{16} \mathbb{E}_4^2D_q^{(1)}-\frac{6825}{8} \mathbb{E}_6D_q^{(2)}-\frac{221375 \mathbb{E}_4 \mathbb{E}_6}{32}\Big)\cZ_{(A_2,A_4)}=0~.
  \ee
  
 {\bf    $(A_2, A_{6  })$ AD theory. }   A monic MLDE  has been found to be 
   \beqn
  &&
 \Bigg[ D_q^{(10)}-\frac{6928}{5} \mathbb{E}_4D_q^{(8)}+115066 \mathbb{E}_4^2D_q^{(6)}-\frac{65828}{5} \mathbb{E}_6D_q^{(7)}+\frac{27134436}{25} \left(\mathbb{E}_4 \mathbb{E}_6\right)D_q^{(5)}
  -\frac{5064230556}{125} \left(\mathbb{E}_4^2 \mathbb{E}_6\right)D_q^{(3)}
   \nonumber   \\&&
        +\left(\frac{4157344488 \mathbb{E}_4^3}{125}-\frac{1904916944 \mathbb{E}_6^2}{25}\right)D_q^{(4)}
    +\left(-\frac{153425310291 \mathbb{E}_4^4}{625}-\frac{48280541232}{125} \mathbb{E}_4 \mathbb{E}_6^2\right)D_q^{(2)}
   \nonumber    \\&&+\left(-\frac{223567701588}{125} \mathbb{E}_4^3 \mathbb{E}_6-\frac{1342610519712 \mathbb{E}_6^3}{125}\right)D_q^{(1)}
      -\frac{6298771946544}{125} \mathbb{E}_4^2 \mathbb{E}_6^2
      \Bigg]\cZ_{(A_2,A_6)}=0~.
  \eeqn

  Meanwhile,   a non-monic MLDE is also found
  \beqn
  & 
 \Bigg[ &
  \mathbb{E}_4D_q^{(8)}-\frac{2128}{5} \mathbb{E}_4^2D_q^{(6)}-28 \mathbb{E}_6D_q^{(7)}+588 \left(\mathbb{E}_4 \mathbb{E}_6\right)D_q^{(5)}-\frac{472164}{25} \left(\mathbb{E}_4^2 \mathbb{E}_6\right)D_q^{(3)}
       \nonumber    \\&&
       +\left(-48230 \mathbb{E}_4^3+\frac{751856 \mathbb{E}_6^2}{5}\right)D_q^{(4)}
       +\left(-\frac{11004312 \mathbb{E}_4^4}{125}-\frac{1880032}{5} \mathbb{E}_4 \mathbb{E}_6^2\right)D_q^{(2)}
              \nonumber    \\&&+\left(-\frac{1370396412}{125} \mathbb{E}_4^3 \mathbb{E}_6+\frac{448292768  }{25}\mathbb{E}_6^3\right)D_q^{(1)}-\frac{8034147891 \mathbb{E}_4^5}{625}
                    \Bigg]\cZ_{(A_2,A_6)}=0~.
  \eeqn

 \subsection{$D_p(SU(N))$  }\label{MLDEADDpSUN}

 We now consider the family of   $D_p(SU(N))$ AD theory   reviewed in \eqref{DpSUNAD}. Here we again impose    the condition $\gcd(p,N)=1$.
 
The central charge of the corresponding chiral algebra is 
\be
c^{D_p(SU(N))}_{2d}=-12c^{D_p(SU(N))}_{4d}=- {(p-1) (N^2-1)} 
\ee

The Schur index is given in \eqref{IDpSUNind}. 
For simplicity, we turn off all the flavor fugacities. Then the corresponding Schur partition function is 
\be
\cZ_{D_p(SU(N))}(q) =q^{-\frac{c^{D_p(SU(N))}_{2d}}{24}}\cI_{D_p(SU(N))}(q, \bm x=1)
=q^{(p-1) (N^2-1)/24}\PE\Big[\frac{(q-q^p)(N^2-1)} {(1-q)(1-q^p)}  \Big]~.
\ee
For   $D_p(SU(N))$  theory, the chiral algebra  is given by the Kac-Moody algebra $\widehat{\frak {su}(N)}_{-\frac{N(p-1)}{p}}$, and the only strong generators are Kac-Moody currents $J^a$ with $a=1,2,\cdots, N^2-1$. In particular, the stress tensor is given via Sugawara construction: 
\be
T\sim \sum_{a }J^a J^a~, \qquad L_{-2}\Omega+\# \sum_{a }  J^a_{-1}J^a_{-1}\Omega=0~.
\ee
But this kind of null relation does not lead to  a MLDE as $L_{-2}\Omega \not\in C_2(\cV)$. 

In case $N=2$, we have AD theories which  can be denoted alternatively as $D_{2k-1}(SU(2))=(I_{2,2k-3},F)=(A_1, D_{2k-1})$. In \cite{Beem:2017ooy}, the authors found  the null state of the form (for small $k$ explicitly) 
\be
J_{-1}^A  (J_{-1}^1J_{-1}^1+J_{-1}^2J_{-1}^2+J_{-1}^3J_{-1}^3)^{k-1}\Omega\in C_2(\cV) ~. 
\ee
As a result $( L_{-2})^k\Omega\in C_2(\cV)$, which leads the MLDE  of weight $2k$.
\footnote{However,   this seems to be not true for  $k=8$. In this case, we find a  MLDE of weight 12 satisfied by the Schur partition function 
$
\Big[ D_q^{(6)}-\frac{3493}{5} \mathbb{E}_4D_q^{(4)}
+\frac{7028}{5} \mathbb{E}_6D_q^{(3)}
+\frac{354331}{25} \mathbb{E}_4^2D_q^{(2)}
+\frac{323204}{5}  \mathbb{E}_4 \mathbb{E}_6 \,D_q^{(1)} 
+\frac{1}{25} \left(6596205 \mathbb{E}_4^3-13130040 \mathbb{E}_6^2\right)
\Big]\cZ_{D_{15}(SU(2))}=0~.$
 }

       {\bf        $D_3(SU(2))$ AD theory. } 
 
      \be
      \Big(D_q^{(2)}-15 \bE_4\Big)\cZ_{D_3(SU(2))}=0~.
      \ee

        {\bf        $D_5(SU(2))$ AD theory. }    
    \be
  \Big(  D_q^{(3)}-\frac{236}{5} \mathbb{E}_4D_q^{(1)}-\frac{756 \mathbb{E}_6}{5}\Big)\cZ_{D_5(SU(2))}=0~.
    \ee

         {\bf        $D_7(SU(2))$ AD theory. }    
    \be
  \Big( D_q^{(4)}-\frac{730}{7} \mathbb{E}_4D_q^{(2)}-\frac{36980}{49} \mathbb{E}_6D_q^{(1)}-\frac{164025 \mathbb{E}_4^2}{343}\Big)\cZ_{D_7(SU(2))}=0~.
    \ee
   
    {\bf         $D_2(SU(3))$ AD theory. } 
      \be
      \Big(D_q^{(2)}-40 \bE_4\Big)\cZ_{D_2(SU(3))}=0~.
      \ee

    {\bf         $D_2(SU(5))$ AD theory. } 
      \be
      \Big(D_q^{(3)}-400 \bE_4D_q^{(1)}\Big)\cZ_{D_2(SU(5))}=0~.
      \ee
 
     {\bf         $D_2(SU(7))$ AD theory. } 
      \be
      \Big(D_q^{(3)}-1840 \mathbb{E}_4D_q^{(1)}+30240 \mathbb{E}_6\Big)\cZ_{D_2(SU(7))}=0~.
      \ee
 
    {\bf         $D_2(SU(9))$ AD theory. } 
      \be
      \Big(D_q^{(3)}-5440 \mathbb{E}_4D_q^{(1)}+196000 \mathbb{E}_6\Big)\cZ_{D_2(SU(9))}=0~.
      \ee
 
   {\bf         $D_2(SU(2k+1))$ AD theory. }   For $p=2$ and $N=2k+1>3$, the above examples suggest that  the corresponding MLDEs  always have  weight 6. Assuming this and using $\cI_{D_2(SU(2k+1))}=1 + 4 k (1 + k) q+\cdots$, we can   determine the general form of MLDE completely
      \be
      \Big(D_q^{(3)}-5 \left(3 k^4+6 k^3-3 k^2-6 k+8\right) \mathbb{E}_4D_q^{(1)}
      +35 \left(k^6+3 k^5-3 k^4-11 k^3-6 k^2\right) \mathbb{E}_6\Big)\cZ_{D_2(SU(9))}=0~.
      \ee
 This suggests that $( L_{-2})^3\Omega \in C_2(\cV)$.
 
\subsection{$\cT_{(3,2)}$ } \label{MLDET23}
The chiral algebra of $\cT_{(3,2)}$ AD theory is given by the $\mathcal A(6)$ algebra    \cite{Buican:2020moo,feigin2007fermionic,feigin2008characters}.
It contains 3 strong generators, denoted by $T, \Phi, \widetilde\Phi$, whose conformal dimensions are 2, 4, 4, respectively. While the first generator $T$ is the stress tensor, the latter two $\Phi, \widetilde\Phi$ are fermionic Virasoro primary operators.  The OPEs among them are given by
\beqn
T(z)T (0)  &\sim& \frac{-12}{z^4}+\frac{2T}{z^2}+\frac{T'}{z} ~,\\
T(z)\Psi (0)  &\sim&  \frac{4\Psi}{z^2}+\frac{\Psi'}{z} ~, \\
T(z)\tilde\Psi (0)  &\sim& \frac{4\tilde\Psi}{z^2}+\frac{\tilde\Psi'}{z}  ~,\\
\Psi(z)\tilde\Psi (0)  &\sim&
-\frac{6 }{z^8} +\frac{2 T}{z^6} + \frac{T'}{z^5} +\frac{3 \left(T''-T^2\right)}{7 z^4}+\frac{2 T^{(3)}-9 T' T}{21 z^3}
\nonumber\\&&
+\frac{-48 \left(T'\right)^2-84 T ''T+36 T^3+7 T^{(4)}}{420 z^2}+\frac{60 \left(-5 T'' T'+6T' T^2  -2 T^{(3)} T\right)+7 T^{(5)}}{2800 z} ~, \qquad\qquad
\eeqn
where we ignore the argument of operators on the RHS, which is 0.
Actually the full OPE can be easily bootstrapped using the associativity of OPEs and the information of   conformal dimensions of these operators.  \footnote{When computing the OPEs, we make heavy use of the Mathematica package\texttt{    OPEdefs} \cite{Thielemans:1991uw}.}

This algebra admits a free field  realization in terms of chiral boson $\varphi$ satisfying the OPE
\be
\varphi(z) \varphi(0)\sim \log z~.
\ee
We can write the generator of the chiral algebra using $\varphi$  \cite{feigin2007fermionic,feigin2008characters}
\beqn
\Psi(w)&=&e^{-\sqrt 3 \varphi(w)}~, \qquad T=\frac12( \p\varphi)^2+\frac{5}{2\sqrt{3}}\p^2\varphi~,  \\
\widetilde\Psi(w) &=&\frac{1}{2\pi i}\oint  \frac{dz}{z-w}e^{2\sqrt{3}\varphi(z)}\Psi(w)
=P_5(\p\varphi, \p^2\varphi, \cdots  \p^5\varphi) e^{ \sqrt 3\varphi(z)} ~,
\eeqn
where $P_5$ is an operator with conformal dimension 5 built as a polynomial in $\p^j\varphi$. Since it is quite complicated, we don't write it down explicitly here. One can verify that the free field realization is consistent with the OPEs above.

The mode expansion of these operators are
 \be 
 T(z)   
  =\sum_{n\in \bZ}L_{n}z^{-2-n}~,\qquad
   \Phi(z)   
  =\sum_{n\in \bZ}\Phi_{n}z^{-4-n}~,\qquad
   \widetilde\Phi(z)   
  =\sum_{n\in \bZ}\widetilde\Phi_{n}z^{-4-n}~.
 \ee
Using the previous OPEs, we can derive the commutation relation of these modes \footnote{In deriving the mode algebra, we   use the  formulae in \eqref{dermode}-\eqref{LOcom}.
}
\beqn
[L_m,L_n]&=& (m-n)L_{m+n}-2(m^3-m)\delta_{m+n,0}~,
\\ 
{} [L_m,\Phi_n]&=& (3m-n)\Phi_{m+n} ~,
\\
{} [L_m,\widetilde\Phi_n]&=& (3m-n)\widetilde\Phi_{m+n} ~,
\\
\{\Psi_m, \widetilde\Psi_n\}&=&
\frac{1}{840} n \left(n^2 \left(n^2-7\right)^2-36\right) \delta _{0,m+n}-\frac{1}{84} (m-n) \left(m^2-m n+n^2-7\right) \Lambda_ { m+n}
\nonumber\\&&
+\frac{1}{1680}(m-n) \Big(3 (m+n)^4-14 m n \left(m^2+m n+n^2\right) -39 (m+n)^2+98 m n+108 \Big) L_{ m+n} 
\nonumber\\&&
-\frac{5}{112} (m-n) \widetilde\Lambda_ { m+n}+\frac{7}{80} (m-n) \Upsilon_{m+n}~,
\label{Phicomm}
\eeqn
 where  \footnote{The $\Lambda,\widetilde\Lambda,\Upsilon$ constructed in this way have   simple commutation relations with stress tensor. E.g. $ [L_m,\Lambda_n]=\frac{1}{30} (5 c+22) m \left(m^2-1\right)L_{m+n} +(3m-n) \Lambda_{m+n}$. They become the null operators of Virasoro algebra for specific values of central charge.  
}
 \beqn
\Lambda&=&T^2-\frac{3}{10}T''~,
\\  
\widetilde\Lambda&=&T^3 -\frac{1}{3} (T')^2 -\frac{19}{30}T'' T-\frac{1}{36}T''''~, \qquad
\Upsilon=T^3 -\frac17 (T')^2 -\frac{11}{14}T'' T-\frac{19}{588} T''''~.
\eeqn

We need to understand the null operators of this VOA. At dimension 6 we find 2 null operators
\be\label{nul6}
 N_6=\Psi''-6T\Psi = 0~, \qquad \widetilde N_6= \widetilde \Psi''-6T\widetilde\Psi= 0~.
\ee
Besides, we also have the null operators  $7 T\Psi-\Psi T=0$ and $7 T\widetilde\Psi-\widetilde\Psi T=0$, which arise from exchanging the order of two operators.

At dimension 8 we find a non-trivial null operator  \footnote{There are of course other null operators which are obtained from the one of lower dimension by taking derivative and multiplying with other operators. They can be obtained straightforwardly, so we will not show them explicitly. }

\be
 N_8=T^4+\frac{140 }{3} \Psi  \widetilde\Psi -\frac{10}{3}   \left(T'\right)^2 T -\frac{10  }{3}   T'' T^2+\frac{5  }{4}\left(T''\right)^2+\frac{10}{9}  T^{(3)} T' +\frac{5}{18}  T^{(4)}T-\frac{5  }{216}T^{(6)}~.
\ee
Due to the presence of $ \Psi  \widetilde\Psi $, the corresponding state does not fit into the form of \eqref{nullT}, namely $(L_{-2})^4 \Omega\not \in C_2(\cV)$. However, we can consider the normal order product   $T N_8$, then  the fermion bilinear terms can be   rewritten $ T\Psi  \tilde\Psi \to  \Psi''  \tilde\Psi$ using the null operator \eqref{nul6}. Consequently, we get a null operator at level 10:
\beqn
 N_{10}&=&
T^5+\frac{4}{3}   T^{(3)} T' T-\frac{10}{3}   \left(T'\right)^2 T^2 -\frac{10    }{3} T''T^3+2  T'' \left(T'\right)^2 +\frac{9}{4}  \left(T''\right)^2T
 -\frac{1}{9} \left(T^{(3)}\right)^2
\nonumber \\&&
 -\frac{1}{9}   T^{(4)} T^2 -\frac{1}{12}  T^{(4)} T''-\frac{1}{30}  T^{(5)}  T'-\frac{13}{360}   T^{(6)}T+\frac{ 1}{5040}T^{(8)}
+\frac{70  }{9}  \Psi ''\tilde{\Psi } +\frac{140 }{3}   \tilde{\Psi }''  \Psi~.
\qquad\quad
\eeqn
 
 In terms of modes, the corresponding null state is
  \beqn
\cN_{10} &=&
 \Big( L_{-2}^5-\frac{20}{3} L_{-4} L_{-2}^3-\frac{10}{3} L_{-3}^2 L_{-2}^2-\frac{8}{3} L_{-6} L_{-2}^2+9 L_{-4}^2 L_{-2}-26 L_{-8} L_{-2}+8 L_{-5} L_{-3} L_{-2}
\nonumber \\ &&\;\;
 -4 L_{-5}^2+4 L_{-4} L_{-3}^2+8 L_{-10}-4 L_{-6} L_{-4}-4 L_{-7} L_{-3}+\frac{280}{3}  \widetilde  \Psi   _{-6}  \Psi   _{-4}+\frac{140}{9} \Psi   _{-6} \widetilde \Psi   _{-4} 
 \Big)\Omega ~.
 \qquad \qquad
  \eeqn
This null state  has the form of \eqref{nullT}, namely $(L_{-2})^5 \Omega  \in C_2(\cV)$. The presence of such a null operator enables us to derive the MLDE. The derivation of   Virasoro mode contribution is the same as before, so we only show the nontrivial contribution from fermionic modes.
  
Using Zhu's recursion relation    \eqref{Zhu2}, we find the insertion of fermion bilinear zero mode  reduces to
  \be
  \Psi   _{-6} \widetilde \Psi   _{-4}  \Omega \quad \to \quad
  \Psi   _{-2+2r} \widetilde \Psi   _{-4} \Omega= \{ \Psi   _{-2+2r} ,\widetilde \Psi   _{-4} \} \Omega~, \qquad r=0,1,2,3,\cdots~,
  \ee
  where we used the property of vacuum that $\Psi_{n}\Omega=\widetilde\Psi_{n}\Omega=0$ for $n>-4$. The  anti-commutators of fermions are given by Virasoro modes in \eqref{Phicomm}. Explicitly, we find 
  \beqn
  \Psi _{-2} \widetilde{\Psi }_{-4}\Omega &=&\Big( \frac{3  }{35}L_{-2}^3-\frac{2}{5} L_{-4} L_{-2}+\frac{2 }{5} L_{-6}-\frac{4  }{35}L_{-3}^2\Big) \Omega~,
  \\
  \Psi _0 \widetilde{\Psi }_{-4} \Omega&=& \Big( \frac{6  }{7}L_{-4}-\frac{3  }{7}L_{-2}^2 \Big) \Omega~,
  \\
  \Psi _2 \widetilde{\Psi }_{-4}\Omega&=& 2 L_{-2}\Omega~,
  \\
  \Psi _4 \widetilde{\Psi }_{-4}\Omega&=& -6\Omega~,
    \\
  \Psi _n \widetilde{\Psi }_{-4}\Omega&=& 0~, \qquad n>4~.
  \eeqn
Therefore, everything reduces to the stress tensor insertion again, which can be computed easily. 
The contribution from inserting the zero mode of $ \widetilde \Psi   _{-6}   \Psi   _{-4}  \Omega $ can computed in exactly the same way; actually their contribution   is just  the opposite  of inserting $   \Psi   _{-6}  \widetilde \Psi   _{-4}  \Omega $, as one can see from the commutation relation  \eqref{Phicomm}.
After combining all contributions together, we arrive at the following simple MLDE
  
  \be\label{MDeqT32}
 \Big[ D_q^{(5)} -140 \mathbb E_{4}   D_q^{(3)} -700 \mathbb E_{6}   D_q^{(2)}-2000   \mathbb E_{4}^2   D_q^{(1)} \Big]
   \cZ_{\cT_{ (3,2)}}=   0~,
    \ee
    where the Schur partition function
  \be 
 \cZ_{\cT_{ (3,2)}}=q^{-c_{2d}/24} \cI
     =  q^{ c_{4d}/2} \cI~, \qquad \cI=1+\cdots~.
  \ee
  Explicitly,    it is given in \eqref{SchT32}.
  One can then numerically verify that the MLDE \eqref{MDeqT32} is indeed satisfied.
    
  \subsection{$\cT_{(4,3)}$  }\label{MLDET43}
  The Schur partition function of $\cT_{(4,3)}$ AD theory is annihilated by the following MLDO of weight 34: 
  
\begin{eqnarray}
  \cD_q^{(17)}&=&
  D_q^{(17)}-\frac{51533242520305}{14352492173} \mathbb{E}_4D_q^{(15)}
  -\frac{6973916257268590}{14352492173} \mathbb{E}_6D_q^{(14)} 
    -\frac{85177879156262050}{14352492173} \mathbb{E}_4^2D_q^{(13)}
 \nonumber  \\ &&
  +\frac{14443540566243470150}{14352492173}  \mathbb{E}_4 \mathbb{E}_6 D_q^{(12)} 
  \nonumber  \\ &&
  +\left(\frac{181610660884391930500 \mathbb{E}_4^3}{14352492173}+\frac{509095491706601844275 \mathbb{E}_6^2}{14352492173}\right)D_q^{(11)}
\nonumber  \\&&
  +\frac{1659222240987437280000}{14352492173}  \mathbb{E}_4^2 \mathbb{E}_6 D_q^{(10)}
\nonumber    \\ &&
    +\left(\frac{15968624650880199775000 \mathbb{E}_4^4}{14352492173}-\frac{381606963489542603408750 \mathbb{E}_4 \mathbb{E}_6^2}{14352492173}\right)D_q^{(9)}
  \nonumber   \\ &&
        +\left(-\frac{2303515395426898055375000 \mathbb{E}_4^3 \mathbb{E}_6}{14352492173}-\frac{3508816208531766930691250 \mathbb{E}_6^3}{14352492173}\right)D_q^{(8)}
      \nonumber     \\ &&
      +\left(-\frac{20430370238300389835000000 \mathbb{E}_4^5}{14352492173}-\frac{22927476579325554250000 \mathbb{E}_4^2 \mathbb{E}_6^2}{161263957}\right)D_q^{(7)}
      \nonumber     \\ &&
          +\left(\frac{33459126792888865915000000 \mathbb{E}_4^4 \mathbb{E}_6}{14352492173}-\frac{333254744268135740656550000 \mathbb{E}_4 \mathbb{E}_6^3}{14352492173}\right)D_q^{(6)}
         \nonumber        \\ &&
              + \Bigg(\frac{350540943451154796600000000 \mathbb{E}_4^6}{14352492173}+\frac{2606434059401166876476000000 \mathbb{E}_4^3 \mathbb{E}_6^2}{14352492173}
         \nonumber       \\ &&      \qquad         
                -\frac{5593009986155179293484100000 \mathbb{E}_6^4}{14352492173} \Bigg)D_q^{(5)}
     \nonumber    \\ &&
            +\left(-\frac{22112801167551810142560000000 \mathbb{E}_4^5 \mathbb{E}_6}{14352492173}+\frac{86682055162600283950200000000 \mathbb{E}_4^2 \mathbb{E}_6^3}{14352492173}\right)D_q^{(4)}
           \nonumber     \\ &&
                +\Bigg(\frac{4244110332161599549800000000 \mathbb{E}_4^7}{14352492173}+\frac{22109643087416294060700000000 \mathbb{E}_4^4 \mathbb{E}_6^2}{14352492173}
      \nonumber    \\ &&      \qquad         
                          +\frac{163123204241712899740202500000 \mathbb{E}_4 \mathbb{E}_6^4}{14352492173}\Bigg)D_q^{(3)}            
                          \nonumber                        \\ &&
                 +\Bigg(\frac{191602959513327117390600000000 \mathbb{E}_4^6 \mathbb{E}_6}{14352492173}+\frac{1734728112424294803141900000000 \mathbb{E}_4^3 \mathbb{E}_6^3}{14352492173}
  \nonumber   \\ &&      \qquad         
                   -\frac{986866725033974720394987500000 \mathbb{E}_6^5}{14352492173}\Bigg)D_q^{(2)}
           \nonumber       \\ &&
                 +\Bigg(-\frac{84640089235700428848000000000 \mathbb{E}_4^8}{14352492173}+\frac{8660372938971876516945600000000 \mathbb{E}_4^5 \mathbb{E}_6^2}{14352492173}
            \nonumber                  \\ &&      \qquad         
                                -\frac{5201608991349440962164600000000 \mathbb{E}_4^2 \mathbb{E}_6^4}{14352492173}\Bigg)D_q^{(1)}~.
                                \label{MLDOT43}
                                \end{eqnarray}

 \bibliographystyle{JHEP} 
 
\bibliography{ref.bib} 
  
\end{document}